\documentclass[final,11pt,3p,authoryear]{elsarticle}

\usepackage{amssymb}
\usepackage[usenames,dvipsnames]{color}
\usepackage{enumitem}

\def\degree{$^\circ$}

\journal{Icarus}

\begin{document}

\begin{frontmatter}



\title{Evidence of Titan's Climate History from Evaporite Distribution}


\author[1]{Shannon M. MacKenzie\corref{cor1}}
\ead{mack3108@vandals.uidaho.edu}
\cortext[cor1]{Corresponding author}

\author[1]{Jason W. Barnes}
 \author[2]{Christophe Sotin}
 \author[3]{Jason M. Soderblom}
  \author[4]{St\'ephane Le Mou\'elic}
  \author[5]{Sebastien Rodriguez}
  \author[6]{Kevin H. Baines}
 \author[2]{Bonnie J. Buratti}
  \author[7]{Roger N. Clark}
 \author[8]{Phillip D. Nicholson}
 \author[9]{Thomas B. McCord}
 
  \address[1]{Department of Physics, University of Idaho, Moscow, ID 83844-0903 USA}
  \address[2]{Jet Propulsion Laboratory, California Institute of Technology, Pasadena, CA, 91109 USA}
 \address[3]{Department of Earth, Atmospheric and Planetary Sciences, MIT, Cambridge, MA 02139-4307 USA}
 \address[4]{Laboratoire de Plan\'etologie et G\'eodynamique, CNRS UMR6112, Universit\'e de Nantes, France} 
  \address[5]{Laboratoire AIM, Centre d'\~etude de Saclay, DAPNIA/Sap, Centre de l'orme des M\'erisiers, b\^at. 709, 91191 Gif/Yvette Cedex, France} 
 \address[6]{Space Science and Engineering Center, University of Wisconsin-Madison, 1225 West Dayton St., Madison, WI 53706 USA}
 \address[7]{U.S. Geological Survey, Denver, CO, 80225 USA}
  \address[8]{Cornell University, Astronomy Department, Ithaca, NY, USA}
   \address[9]{Bear Fight Institute, P.O. Box 667, 22 Fiddler's Road, Winthrop, WA 98862, USA}

\begin{abstract}
Water-ice-poor, 5-$\mu$m-bright material on Saturn's moon Titan has previously been geomorphologically identified as evaporitic. Here we present a global distribution of the occurrences of the 5-$\mu$m-bright spectral unit, identified with Cassini's Visual Infrared Mapping Spectrometer (VIMS) and examined with RADAR when possible. We explore the possibility that each of these occurrences are evaporite deposits. The 5-$\mu$m-bright material covers 1\% of Titan's surface and is not limited to the poles (the only regions with extensive, long-lived surface liquid). We find the greatest areal concentration to be in the equatorial basins Tui Regio and Hotei Regio. Our interpretations, based on the correlation between 5-$\mu$m-bright material and lakebeds, imply that there was enough liquid present at some time to create the observed 5-$\mu$m-bright material. We address the climate implications surrounding a lack of evaporitic material at the south polar basins: if the south pole basins were filled at some point in the past, then where is the evaporite?
\end{abstract}

\begin{keyword}
Titan \sep Titan surface \sep Spectroscopy \sep Infrared observations \sep Geological processes
\end{keyword}

\end{frontmatter}


\section{Introduction}
\label{intro}
Titan, unique among other satellites in our solar system, has a thick atmosphere in which a volatile (methane, though ethane is thought to also play an important role) precipitates and evaporates in the same fashion as water in Earth's hydrological cycle \citep{Roe2012}. Our understanding of Titan's surface and climate has evolved from the once widely expected global surface ocean thought to sustain the photolytic processes in the atmosphere \citep[e.g.,][]{1983Sci...222.1229L,1983Sci...221...55F} to a world only sparsely covered by liquid deposits but hosting a collection of strikingly Earth-like surface morphologies (e.g., dunes and channels) thanks to data now available from Cassini-Huygens. These observations reveal Titan to be generally wet at the poles and dry at the equator. \\

At the poles, Titan's liquid bodies range from seas (surface area greater than 100,000 km$^2$, only in the north) to smaller lakes \citep{2007Natur.445...61S,Hayes2008,2012Icar..221..768S}, to fluvial features thought to be drainage networks \citep[e.g.,][]{2008P&SS...56.1132L,2008Icar..195..415L, 2009GeoRL..3622203B, 2012P&SS...60...34L, Burr2013morph}. The equatorial region, however, is characterized by expansive dune fields \citep[e.g.][]{2006Natur.441..709E,2007P&SS...55.2025S,2008Icar..195..415L,2008Icar..194..690R,Rodriguez2014}, intermittent mountain chains \citep[e.g.,][]{RADARmountains, CookMtns}, and a noticeable lack of permanent surface liquid (but see \citet{2012Natur.486..237G}). Rain has been observed to wet the midlatitude and near-equatorial surface \citep{2011Sci...331.1414T}, but standing bodies similar in extent and stability to the polar lakes have yet to be conclusively identified. The most equator-ward lake candidates, Sionscaig and Urmia, are located between 39\degree S and 42\degree S and, while hypothesized to have standing liquid, are yet to be confirmed as either long-lived or ephemeral \citep{VixieLakes}. \citet{2008JGRE..113.8015M} were able to reproduce a dry equatorial climate with a global circulation model (GCM) where a limited reservoir of methane exchanges between the atmosphere and surface.\\

And yet, sizable volumes of liquid are still thought to have played an important role in forming the equatorial region. As Huygens discovered during its descent to the Titanian surface on January 14, 2005, fluvial features such as rounded cobbles, channels, and valleys also dot the equatorial landscape \citep{2005Natur.438..765T}. The subsurface structure controlling fluvial drainage networks in southwestern Xanadu\citep{2009GeoRL..3622203B} agrees with the wind flow derived from aeolian driven dune morphology \citep{2009GeoRL..36.3202L}. Rainfall has been proposed as the source the liquid responsible for carving fluvial features \citep{2005P&SS...53..557L,Ontario.dries.up}. While storms have been observed in the equatorial region \citep{2011Sci...331.1414T}, Cassini has yet to directly observe actively flowing channels. The average global rainfall on Titan has been suggested to be $\sim$ 1 cm/yr \citep{1996Icar..122...79L, 2006Sci...311..201R}, but such a low rate could be reconciled with the observed fluvial features if precipitation were to occur in intense but infrequent storms (``methane monsoons" akin to what terrestrial deserts experience though an order of magnitude smaller \citep{Jaumann08,2012Natur.481...58S}). \\

Clouds have been frequently observed in the active atmosphere of Titan, ranging from the enormous winter polar vortex, to bands of tropospheric clouds around certain latitudes \citep[e.g.][]{Griffith09,2009A&ARv..17..105H,2009Natur.459..678R,2010Icar..205..571B,2011Icar..216...89R,2012P&SS...60...86L}, to the low lying fog \citep{2009ApJ...706L.110B}. Stratospheric and high tropospheric clouds have been deduced to be composed of methane and ethane, but fog is necessarily made of methane alone \citep{2009ApJ...706L.110B}. Precipitation has been indirectly observed after cloud coverage via the surface darkening, brightening, and subsequent return to the original spectrum by Cassini's Visual Infrared Mapping Spectrometer (VIMS) \citep{PrecipSurfaceBrightening} and Imaging Science Subsystem (ISS) \citep{2009GeoRL..3602204T,2011Sci...331.1414T,PrecipSurfaceBrightening}. \\

The present distribution of Titan's lakes is asymmetric: there are more in number and extent at the north pole than at the south \citep{2009NatGe...2..851A}. In the north, for example, the largest body is Kraken Mare, a sea covering 400,000 km$^2$ \citep{2009GeoRL..3602204T}, while in the south, the largest is Ontario Lacus which covers only 15,000 km$^2$ \citep{Hayes2010}. Global Circulation Models (GCMs) have found that some kind of seasonal exchange or link between the north and south poles may be taking place  \citep[e.g.,][]{Tokano05,2007Icar..186..385M,2006Sci...311..201R,2009Icar..203..250M,2009Icar..204..619T}. While changes in the size of the northern lakes and seas were not observed before equinox \citep{Hayes2011,2012Icar..221..768S}, Cassini has begun to witness signs of seasonal transport as Titan approaches northern summer. The Composite Infrared Spectrometer (CIRS) has recorded evidence that subsidence has just started above the south pole \citep{2012Natur.491..732T} while the cloud distribution observed by VIMS is indicative of a pole-to-pole meteorological turnover \citep{2011Icar..216...89R,2012P&SS...60...86L}.\\

On a longer than seasonal timescale, it has been proposed that the observed dichotomy actually reverses every $\sim$ 50,000 years with a Titanian Milankovich cycle \citep{2009NatGe...2..851A}. This cycle would be driven by changes in solar insolation as to which hemisphere receives a more intense summer. A GCM by \citet{2012Natur.481...58S} shows that an asymmetry in solar insolation can explain the difference in liquid distribution between the north and the south poles. That work, however, assumes a static amount of methane in the surface-atmosphere system, which may not be the case on Titan, as indicated by other models \citep{2012ApJ...749..159N} and empirical estimates for the age of the atmosphere \citep{2005Natur.438..779N, 2012ApJ...749..160M}. \\

Dry and partially filled lakes have been identified with the Cassini RADAR mapper (RADAR) \citep{Hayes2008, evaporite} both in isolation and near filled lakes. \citet{Hayes2011} also observed ephemeral lakes in the south polar region. These dry beds imply that the distribution of Titan's liquid is not static. \citet{Moore.Tui.Hotei.Lakes} propose that the landscape elements with crenulated margins and small lacustrine features similar to those identified by \citet{Hayes2008,Hayes2011} observed in RADAR images of Tui and Hotei Regiones indicate that the previous presence of liquid, i.e. that the two equatorial features may be fossil seas. Thus, because the small lacustrine features observed within Tui and Hotei would require long-lived surface liquid for their creation, there seems to be evidence for a previous global liquid distribution that differs from what is presently observed.\\

The 5-$\mu$m-bright VIMS spectral unit has been observed on Titan since the first Cassini flybys where Tui Regio is clearly differentiated from other surface features \citep{2005Sci...310...92B}. However, from observations of a region of small lakes and dry lake beds by VIMS during T69 (2010 June 5), \citet{evaporite} established the connection between lakebeds and 5-$\mu$m-bright material: this unique spectral unit coincided with the shores of filled lakes and the bottoms of dry lakes identified by RADAR \citep{Hayes2008}. Within this small region, there are also examples of dry and filled lakes that do not exhibit a 5-$\mu$m-bright signature. \citet{evaporite} demonstrated how evaporite formation would be consistent with these observations of the water-ice poor, uniquely 5-$\mu$m-bright material: the compounds making up the deposits would have re-crystalized upon solute evaporation (accounting for the uniquely bright, water-ice poor spectrum) and the formation process would only occur in saturated solutions (explaining the spectral differences between such geographically close lakes and lakebeds).  While we prefer this evaporitic interpretation for 5-$\mu$m-bright material formation, other, lacustrine-related explanations could be possible, such as sedimentary deposits forming at the bottom of lakebeds. Thus, we discuss the 5-$\mu$m-bright material as ``evaporite candidates" to reinforce the connection between lakebeds and 5-$\mu$m-bright material for ease of comprehension. \\

In this paper, we identify all other examples of the 5-$\mu$m-bright spectral unit in the VIMS data. While brightness at 5 $\mu$m alone is not diagnostic enough to say definitively that an observed signature is from specifically evaporitically formed material, the strength of the geomorphological correlation between lakebeds and this spectral unit (first demonstrated by \citet{evaporite} and bolstered in this work with available RADAR images) gives us enough ground to explore the implications for the global liquid distribution on Titan if each localized 5-$\mu$m-bright signature were indeed indicative of some previous presence of surface liquid. Isolating the specific chemical compounds of evaporite is beyond the scope of this paper. In Section \ref{methods}, we describe the 5-$\mu$m-bright spectral unit, define our candidate selection criteria, and detail the mapping process. Our results are given in Section \ref{results} and their implications on the past climate on Titan are discussed in Section \ref{discussion}. We conclude with a summary of our interpretations, scenarios inspired by the evaporite candidate distribution for GCMs to consider, and propositions for future study.

\section{Methods}
\label{methods}
\subsection{Selection Criteria}

In a hydrological cycle\footnote{Here we refer to any cycle similar to that of water on Earth as ``hydrological" to avoid clumsy and somewhat inaccurate terminology; ``methanological" would seem to belie the role of liquid ethane, for example.}, surface liquid evaporates into the atmosphere where it eventually condenses and rains down onto the surface. There the liquid can dissolve surface material if the solvent is in contact with the solute long enough. For liquid reservoirs that are saturated, the evaporation of the liquid will cause the solute to precipitate out and deposit as evaporite either onto the surface exposed after the liquid has left or at the bottom of the bed of saturated liquid. \\

On Titan, the surface liquid is a predominately methane-ethane mixture \citep{T38.ethane}. It is expected that while the formation processes of evaporite on Titan are similar to those on Earth, the composition will be different because, unlike water, methane and ethane are non-polar. Most of the suggested candidates for compounds that would be soluble in such a mixture are organic \citep{2009ApJ...707L.128C,evaporite,2013Icar..226.1431C}. It could be that the organic compound has a finite vapor pressure such that it could itself evaporate and condense nearby. However, we build upon the geomorphological evidence of \citet{evaporite} and use this 5-$\mu$m-bright spectral unit to identify possible evaporitic deposits.\\

In Figure \ref{spectra}, we show examples of the 5-$\mu$m-bright spectral unit (red line, all panels) in the context of other surface spectral units: the dark brown dunes (brown line, panel a), the bright terrain of Xanadu (green, panel a), the liquid of Kraken Mare (blue, panel b), the non-Xanadu equatorial bright material (light blue, panel b), the dark blue material (dark blue, panel b). We also show the spectral correlation between shoreline evaporite and Hotei Regio (black) in panel c. To compare individual spectra, it is necessary to take into account viewing geometry which necessarily has an effect on the observed signal intensity (as discussed in \citet{SolomonidouPCP}). Spectra of the same region taken near the limb will look different from one at nadir as the former's signal must travel through more atmosphere, for example. Thus, in Figure \ref{spectra}, we group the data based on viewing angles, which are summarized in panel d, along with the ratio of I/F at 2.8 and 2.7 $\mu$m (affected by the presence of pure water ice \citep{Rodriguez.landingsite}). For the behavior of the 5-$\mu$m-bright material in the 5 $\mu$m window, we refer the reader to Section 7.2 and Figure 27 of \citet{McCord2008}, where outlying pixels are compared to the scene average of Tui Regio.  \\

\begin{figure}
 \begin{tabular}{c}
\includegraphics[width=0.7\textwidth]{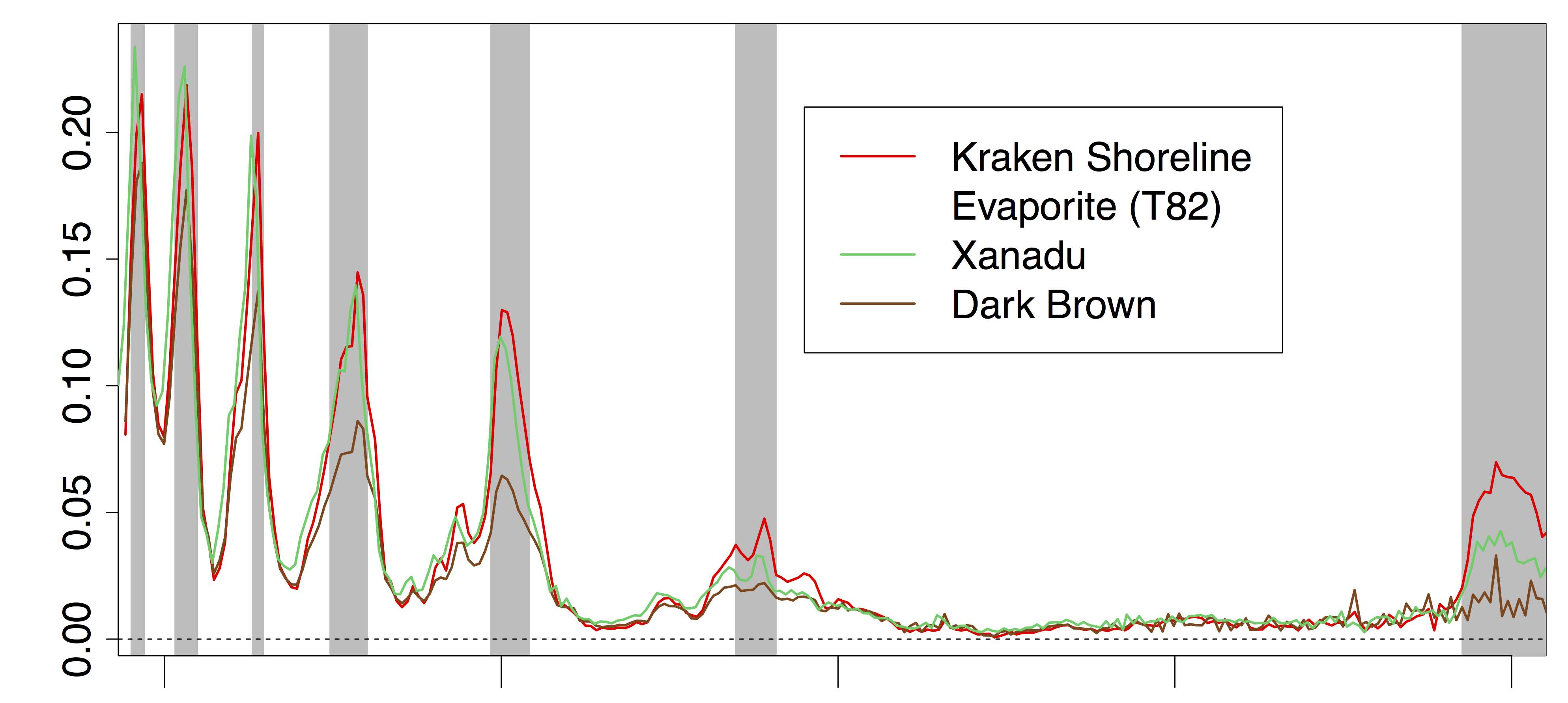}\\
\includegraphics[width=0.7\textwidth]{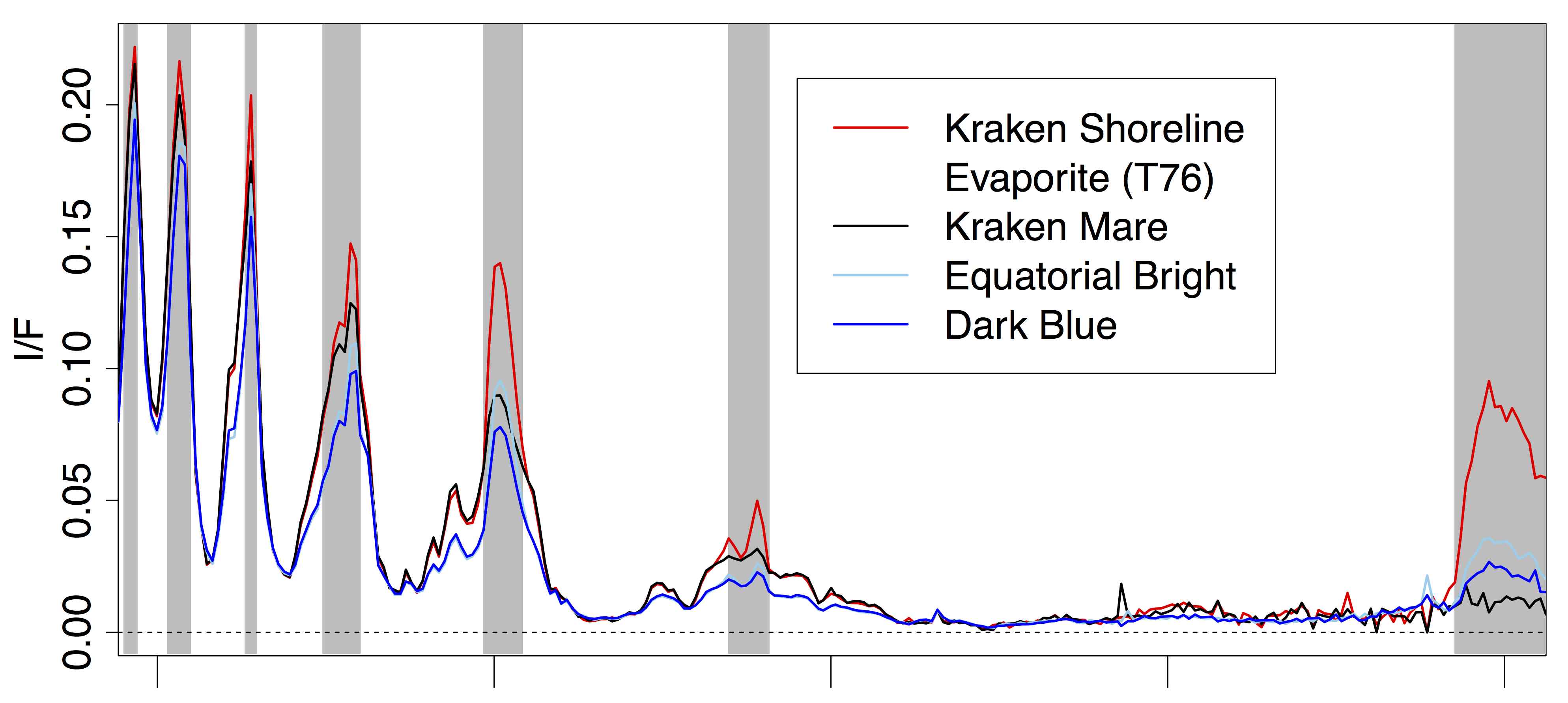}\\
\includegraphics[width=0.7\textwidth]{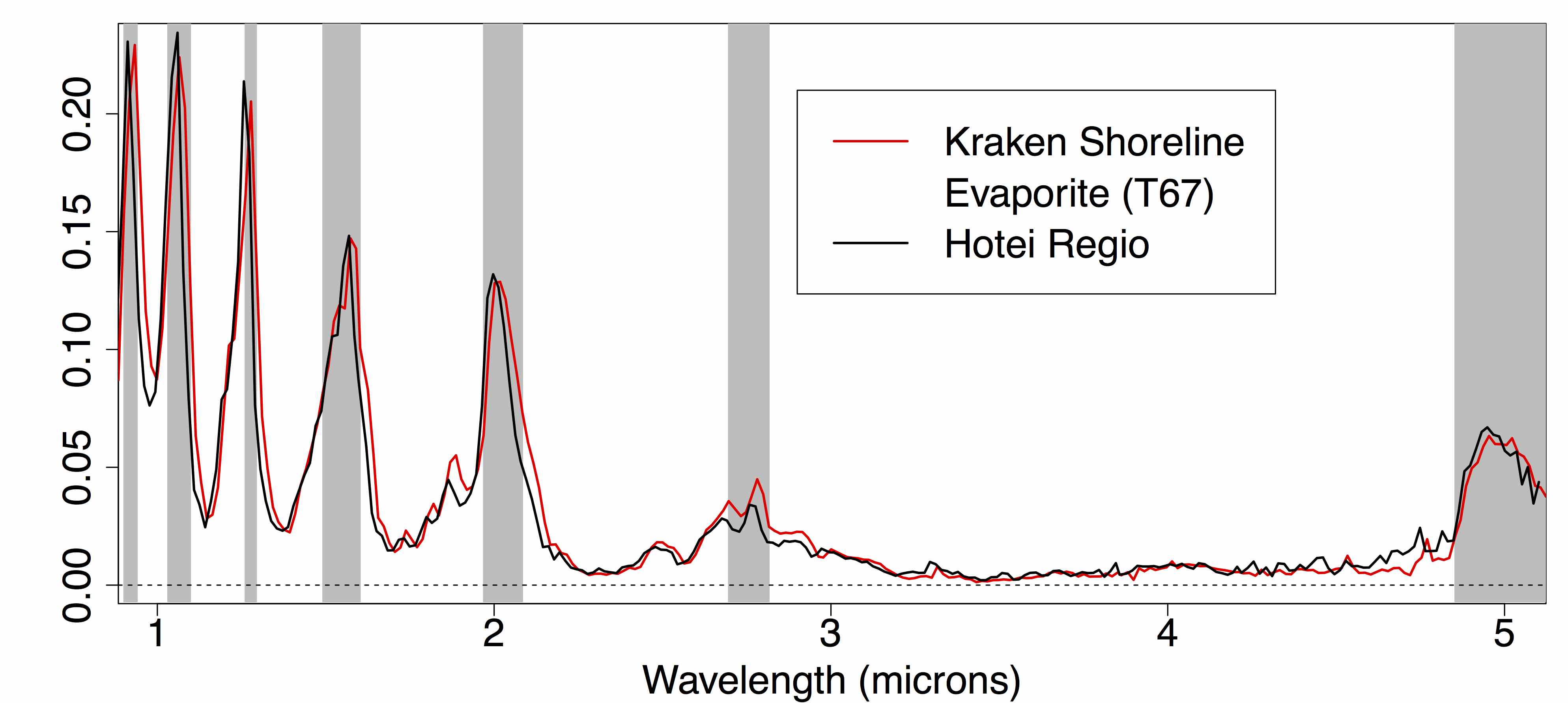}\\
\includegraphics[width=0.7\textwidth]{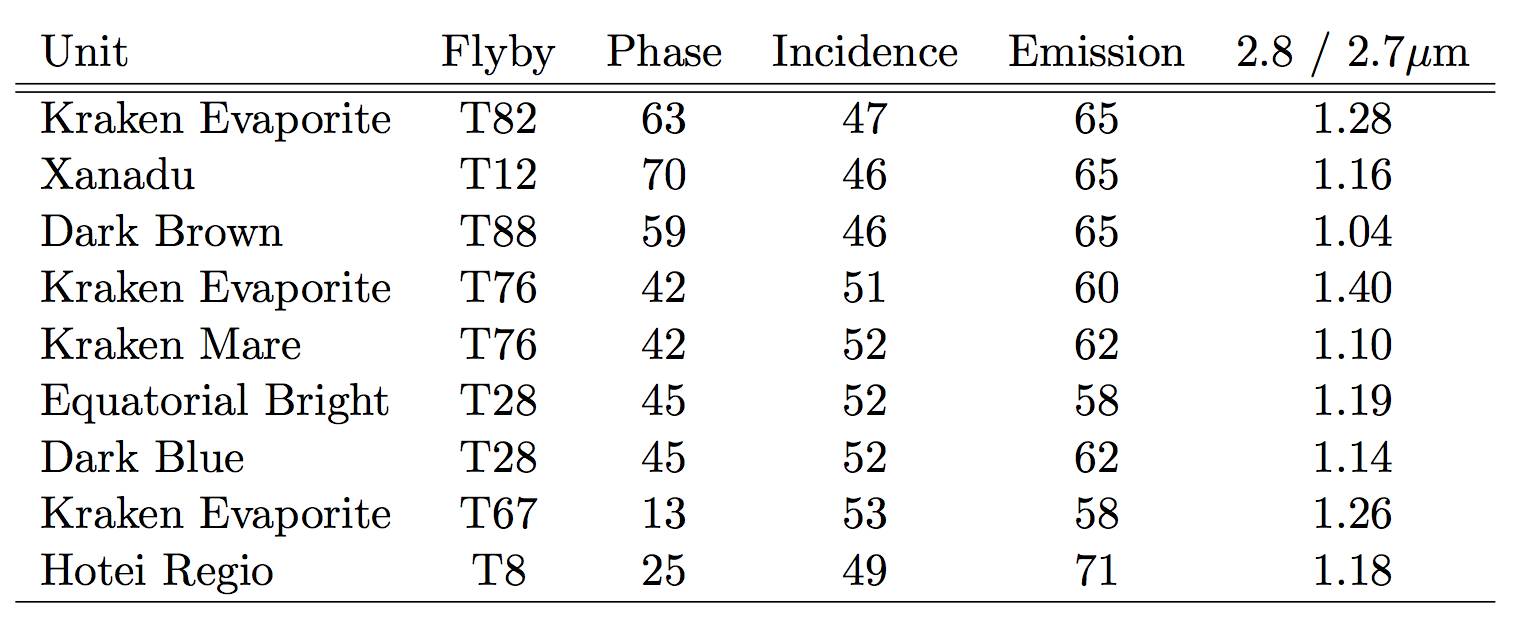}
 \end{tabular}
  \caption{The unique spectrum of 5-$\mu$m-bright material found on the shorelines of Kraken Mare (red in each panel) compared to those of other identified surface feature VIMS spectral units (listed in \citet{2007mapping}): dark brown (brown, panel a), Xanadu (green, panel a), liquid (black, panel b), non-Xanadu equatorial bright (light blue, panel b), dark blue (dark blue, panel b). In panel c, we also show the correlation between shoreline evaporite and the spectral signature of Hotei Regio (black). We have broken up the spectra into different plots based on viewing geometry in order to make appropriate comparisons. The viewing angles and ratio of the I/F at 2.8 and 2.7 $\mu$m is listed in the table of panel d. The grey boxes indicate the wavelength windows in which VIMS can observe Titan's surface. The 5-$\mu$m-bright material is also bright in the 2 micron windows, though not as distinctly as in the 5 $\mu$m window. The spectrum of Kraken Mare is rather bright in T76 which may be due to specular reflections of the sky reflecting off the surface of the liquid \citep{VixieLakes}.} 
\end{figure}

The 5-$\mu$m-bright unit is named after its most dramatic characteristic: material of this composition is brighter than all other spectral units at 5 $\mu$m. However, the titular specular feature is not the unit's only distinction. 5-$\mu$m-bright material has a high ratio of I/F at 2.8/2.7. Generally, in the wavelength windows greater than 2 $\mu$m, the 5-$\mu$m-bright unit is brighter than any other surface unit. The I/F of the 5-$\mu$m-bright unit is on the order of other spectral units (and even darker than Xanadu) in the shortest wavelength windows. Instead of working in the spectral domain, we consider the data spatially by assigning  to the VIMS data a color scheme that exploits the spectral characteristics of the 5-$\mu$m-bright unit. With R = 5 $\mu$m, G = 2 $\mu$m, and B = 1.3 $\mu$m, the evaporite-correlated unit appears as a very bright, reddish-orange color. This method facilitates quick identification of 5-$\mu$m-bright material in VIMS maps, enabling our search to encompass the breadth of available data.\\

Titan's clouds also appear bright at 5 $\mu$m but can be generally distinguished from evaporite. Clouds exhibit a distinctive spectral signature in the wings of the 2 $\mu$m peak due to their altitude \citep{1998Natur.395..575G, 2006Sci...313.1620G, 2005Sci...310..474G,2011Icar..216...89R}, allowing for identification of tropospheric and stratospheric clouds (that is, at least 40 km above Titan's surface). Additionally, clouds demonstrate a simultaneous increase in all windows, particularly evident at the 2.75 $\mu$m and 5 $\mu$m windows \citep{2009Natur.459..678R} and are ephemeral. \\

Low lying vapor, what we broadly refer to as fog, does not demonstrate the 2 $\mu$m wing signature but otherwise has the spectral signature of a cloud. Hence, to identify fog and check the longevity of an evaporite candidate, we compare between VIMS coverage of the same area at different flybys. For each case we ask: Is the 5-$\mu$m-bright material present? Does it change shape and extent? Admittedly, such criteria are less robust than the spectral characteristic of higher altitude clouds. After all, it could be that an evaporitic deposit is no longer seen (or no longer has the same shape) at a different time because the surface has been wetted; the evaporite is either submerged or even dissolved again into the covering liquid. The persistence of an evaporitic deposit is more telling than any ``sudden" disappearance. Low altitude clouds would not statically endure for long timescales, thus the persisting signature would seem to be evaporite. Indeed, if no disappearances were observed over a long enough time scale, for example, one could surmise something about the lack of sufficient rainfall to re-dissolve the evaporite into solution. Realistically, however, such a measurement would require coverage more extensive and more frequent than currently available.  We therefore require that a 5-$\mu$m-bright feature be seen at the same location and of the same shape in at least two flybys in order to be considered an evaporite candidate. The grey polygons of Figure \ref{figure:globaloverview} represent surface features that could not be identified as candidates due to lack or repeat data of sufficient quality. \\ 
\subsection{Mapping}

To identify evaporite candidates, we searched through images taken by VIMS from T0-T94 (July 3, 2004 - January 1, 2014). The spectroscopic data were reduced using the VIMS pipeline detailed in \citet{2007mapping} and geometrically projected with software developed for \citet{T20dunes}. Color images were then created by assigning individual wavelengths the RGB= 5, 2, 1.3 $\mu$m color scheme to capitalize on the unique spectral characteristics of the 5-$\mu$m-bright unit as discussed above. (This color scheme was also used by \citet{2007mapping} and \citet{evaporite}). For topographical and geological context, we use data from the RADAR instrument taken in either Synthetic Aperture RADAR (SAR) or high altitude Synthetic Aperture RADAR (HiSAR) modes that spatially coincides with 5-$\mu$m-bright material observed by VIMS. \\

In Table 1, we list the locations of each evaporite deposit that is mapped in this work, as well as the calculated surface area. Polygons representing the outline of each evaporite candidate are plotted as a global distribution in the bottom half  of Figure \ref{figure:globaloverview}. We also include a cylindrical map of Titan as seen by VIMS in Figure \ref{figure:globaloverview} to provide global context for each candidate's location. From the surface areas, we estimate the volumes of liquid bodies, as well as the hypothetically filled liquid bodies corresponding to evaporite deposits. The depths are estimated from the assumption that depth is proportional to surface area using the measured values for Ligeia Mare \citep{LigeiaDepth}. \\
\begin{table}
\caption{ \small{5-$\mu$m-bright deposits identified as evaporite in order of decreasing latitude. Coordinates are given in the positive-west longitudinal convention. Areas were calculated by summing over the area of each pixel spanned by a polygon outlined with software first designed for \citet{T20dunes}. Data obtained in flybys before T86 are publicly available on the Planetary Data System. The best resolution given below is an average of the best latitudinal and longitudinal resolution available. The listed emission angle is also from the resolution image. Evaporite deposits cover 1\% of the surface of Titan.} }
\tabcolsep=0.1cm
\footnotesize
 \begin{tabular}{l c c c c c c}
\hline
	&	Central 	&	Central &	Total Evaporite 	&	Best VIMS	&	Best & Emission \\
	&	Latitude	&	 Longitude	&	Area	&	Flyby	&	Resolution	& Angle\\
	&		&	(Positive west)	&	(km$^{2}$)	&		&	(km/pixel)	& (\degree)\\
	\hline
	\hline
90 N - 80 N	&		&		&	21,500	&	T90-T94	&	&	\\ 
Punga Mare	&	88	&	150	&	1,220 &	T93, T94	&	6	& 16 - 22\\ 
Kutch	&	88	&	222	&	13,800	&	T93, T94	&	7	& 17 - 22 \\
Ligeia Mare	&	75	&	170	&	4,250	&	T69, T94	&	1	& 14 - 60\\ 
Kivu Lacus	&	87.02	&	118	&	88	&	T85, T93, T94	&	1& 47	\\
Muggel Lacus	&	83	&	170	&	30,000	&	T93, T94	&	7	& 15 - 34\\ 
80 N - 70 N	&		&		&	16,800	&		&		&\\ 
MacKay Lacus	&	77	&	96	&	14,600	&	T94	&	8&	5 - 21 \\ 
Lake District	&	64-83	& 183 - 125 &	55,900	&	T97	&	15&	5\\ 
70 N - 60 N 	&		&		&	18,300	&		&		&\\ 
Cardiel Lacus	&	68	&	203	&	7,100	&	T90	&	62	& 70 \\ 
Woytchugga Lacuna	&	69	&	110	&	66,700	&	T97	&	13	& 9\\ 
Nakuru Lacuna	&	65.5	&	92.38	&	2,580	&	T97	&	13	& 18\\ 
Vanern Lacus	&	69.9	&	223.2	&	7,060	&	T69	&	14	& 63\\ 
Towada Lacus	&	69.3	&	242	&	782	&	T69	&	10	& 63\\
Atacama Lacuna	&	67.6	&	226.1	&	799	&	T69	&	13	& 62\\
Djerid Lacuna	&	65.8	&	219.7	&	808	&	T69	&	13	& 59\\
Uyuni Lacuna	&	65.7	&	223.8	&	700	&	T69	&	13	& 59\\ 
Ngami Lacuna	&	67.2	&	211.2	&	1,020	&	T69	&	13	&60\\ 
60 - 50 	&		&		&	6,020	&		&		&\\ 
Kraken Shores	&	60	&	305	&	32,300	&	T76	&	20	& 73\\ 
South of Kraken	&		&		&	42,000	&	T76	&		&\\ 
50 N - 40 N 	&		&		&	2,810	&	T76	&		&\\ 
Hammar Lacus	&	47	&	313.49	&	18,600	&	T76	&	31	& 55\\
40 N - 30 N	&		&		&	17,200	&		&		&\\
West Fensal 	&	15	&	52	&	18,300	&	T5	&	9	& 22-37\\
East Fensal	&	17.4	&	39.5	&	3,660	&	T5	&	37	& 18\\
West Belet North 	&	-1.5	&	284.9	&	1,160	&	T61	&	15&	48\\ 
West Belet South 	&	-8.2	&	283.9	&	3,650	&	T61	&	15	& 48\\ 
Southwest Belet	&	-18.2	&	276.6	&	3,930	&	T61	&	22	& 57\\ 
Hotei Regio	&	-25.8	&	81.2	&	211,000	&	T48	&	14	& 73\\ 
Tui Regio	&	-25.8	&	122	&	296,000	&	T12	&	43	\\
North Yalaing Terra	&	-15.2	&	322.46	&	2,300	&	T58	&	20	& 16\\	
Ontario Lacus	&	-72	&	180	&	2,610	&	T38, T51	&	1&	28 - 75\\ 
Arrakis Planitia	&	-78.1	&	110.5	&	27,500	&	T51	&	6	&\\	
78 S - 82 S	&		&	&	16,400	&	T23	&		&\\ 
	&		&		\textbf{Total}		&\textbf{998,000}	&	&		&\\
\label{evapcandtable}
 \end{tabular}
\end{table}

\section{Results}
\label{results}

\begin{figure}
 \centering
 \noindent \includegraphics[width=40pc]{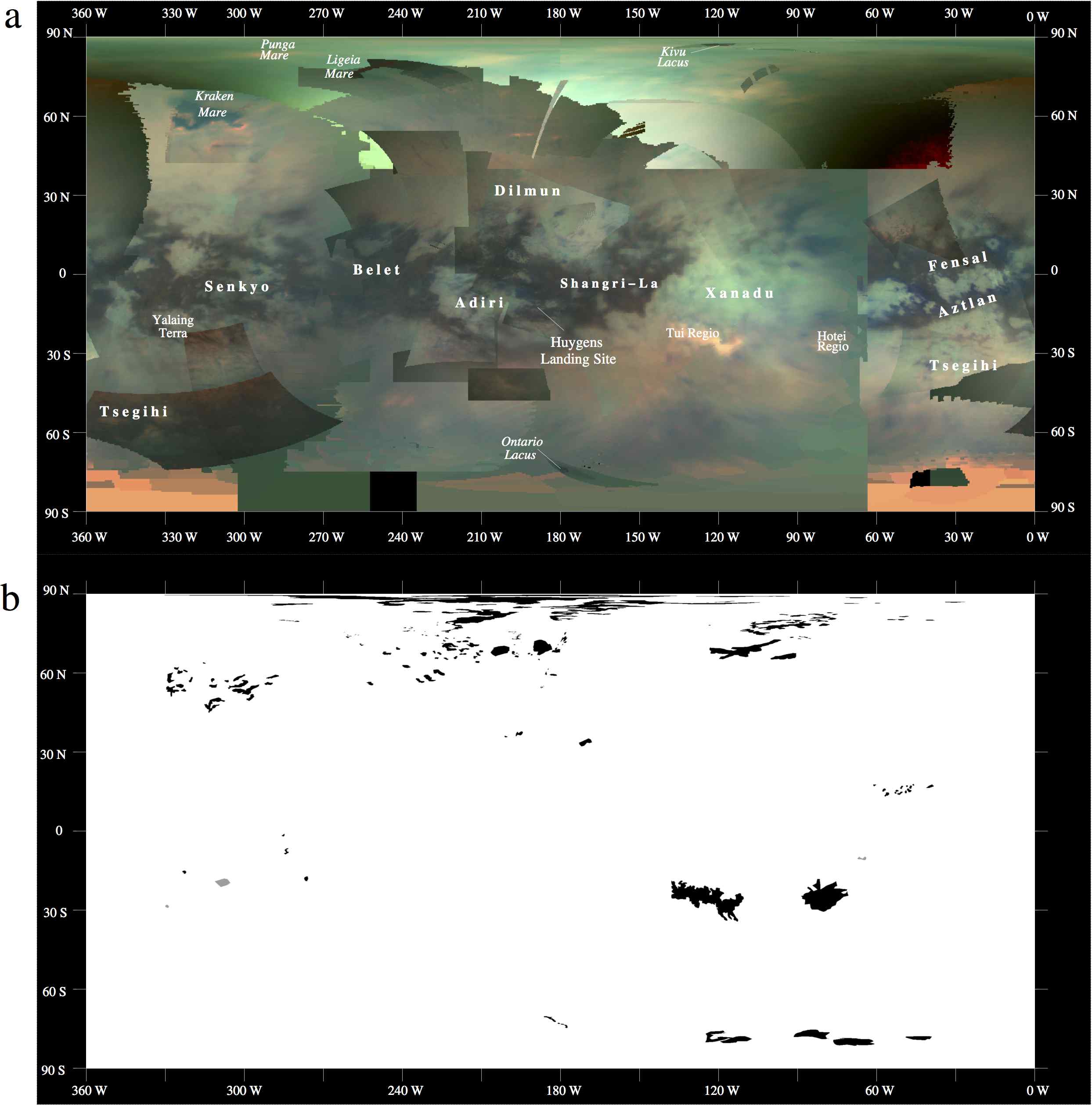}
 \caption{ (a) Global cylindrical map of Titan as seen by Cassini VIMS using the color scheme of \citet{2007mapping}: R= 5 $\mu$m, G = 2 $\mu$m, and B= 1.3 $\mu$m. This VIMS base map uses data from T8-T90. (b) Global distribution of evaporitic deposits in black (listed in Table \ref{evapcandtable}) as well as several 5-$\mu$m-bright areas that did not meet sufficient criteria (e.g. not observed twice, no data of high enough resolution) to be considered evaporite candidates in grey. The polygons shown here, created with software designed for the analysis of \citet{T20dunes}, were used to calculate the area covered by evaporite. While many of the new deposits identified in this study are located near liquid (on or near the shores or empty bottoms of polar lakes and seas), the existence of 5-$\mu$m-bright material in the equatorial region is evidence for the presence of large scale tropical seas in this area sometime in Titan's past. }
\label{figure:globaloverview}
\end{figure}

\begin{figure}
 \noindent\includegraphics[width=20pc]{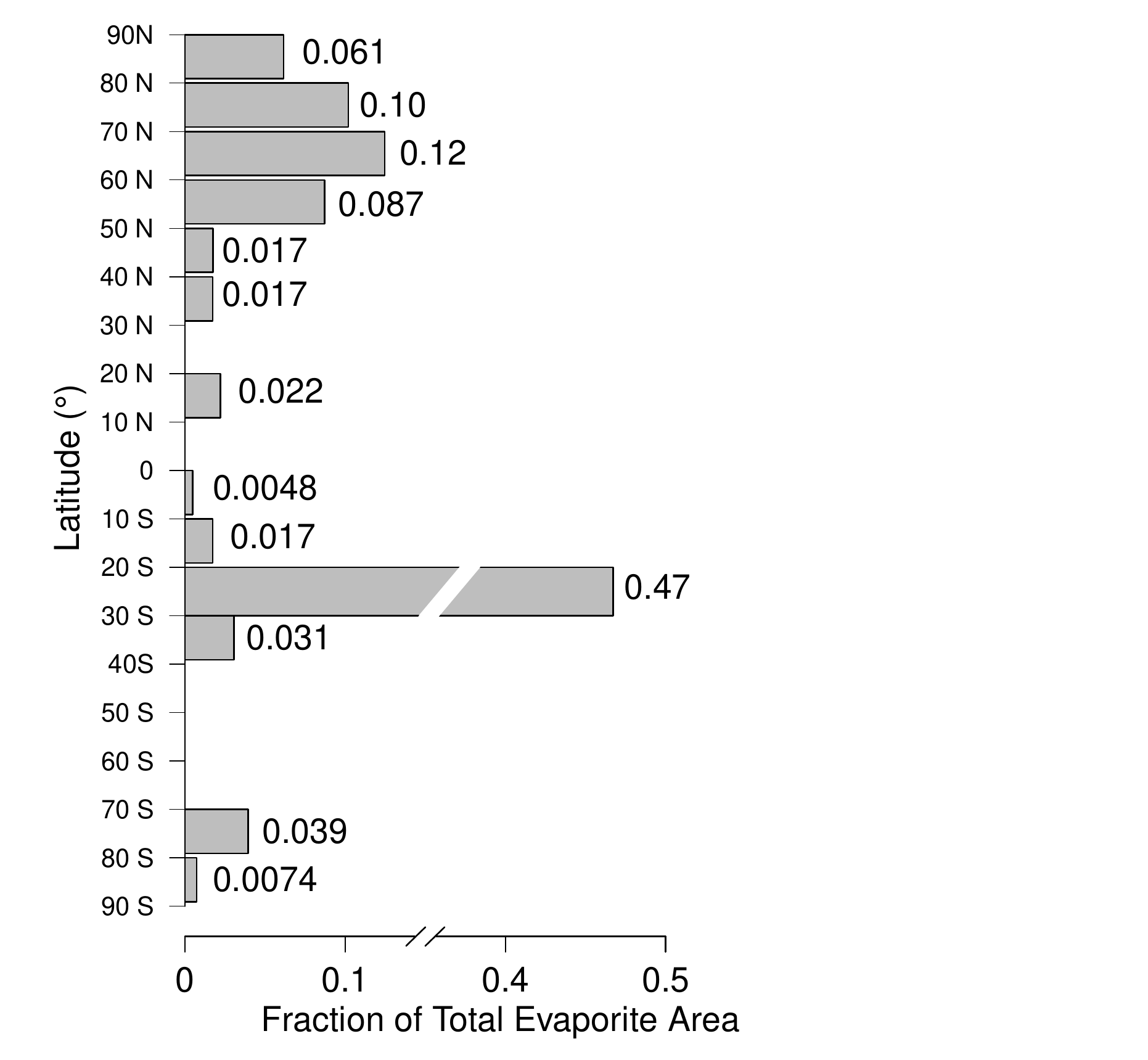}
 \caption{Fraction of total surface area covered by evaporite candidates binned by 10$^{\circ}$ in latitude (fraction listed for each bin by the number to the right of each column). In total, 5-$\mu$m-bright deposits cover a little more than 1$\%$ of Titan's surface. Though the north polar region has the largest amount of known surface liquid, it is the southern equatorial basins Tui Regio and Hotei Regio (located in the 10\degree S - 40\degree S) that contain the most 5-$\mu$m-bright material. The north pole (latitude $>$ 60\degree N) is the region with the next greatest evaporite occurrence. This is unsurprising in light of the formation scenario proposed by \citet{Moore.Tui.Hotei.Lakes}: if Tui Regio and Hotei Regio are fossil sea beds, then all possible compounds precipitated out when the liquid evacuated. The seas of the north pole, the only bodies capable of producing a comparable amount of evaporite, are still covered by liquid and therefore have either not precipitated out the maximum amount of material or some evaporite has fallen to sit on the seafloor. The south pole, notably, is as devoid of evaporite as the northern midlatitudes.}
 \label{figure:fractions}
 \end{figure}

We identify new evaporite candidates in the north polar region, the northern midlatitudes, the equatorial band, and the south polar region, listed in Table \ref{evapcandtable} and shown in Figure \ref{figure:globaloverview}.  If  the connection between 5-$\mu$m-bright material and lakebeds in the north pole holds for other regions, then the areas outlined in Figure \ref{figure:globaloverview} would have been covered with enough liquid at least once to have created the 5-$\mu$m-bright deposits. Interestingly, the largest evaporite candidates are not located in the poles, where the largest amount of surface liquid currently resides, but rather in the equatorial band (Figure \ref{figure:fractions}). The following subsections discuss the deposits in detail by region. 

\subsection{North Pole} 
 \begin{figure}
 \noindent\includegraphics[width=40pc]{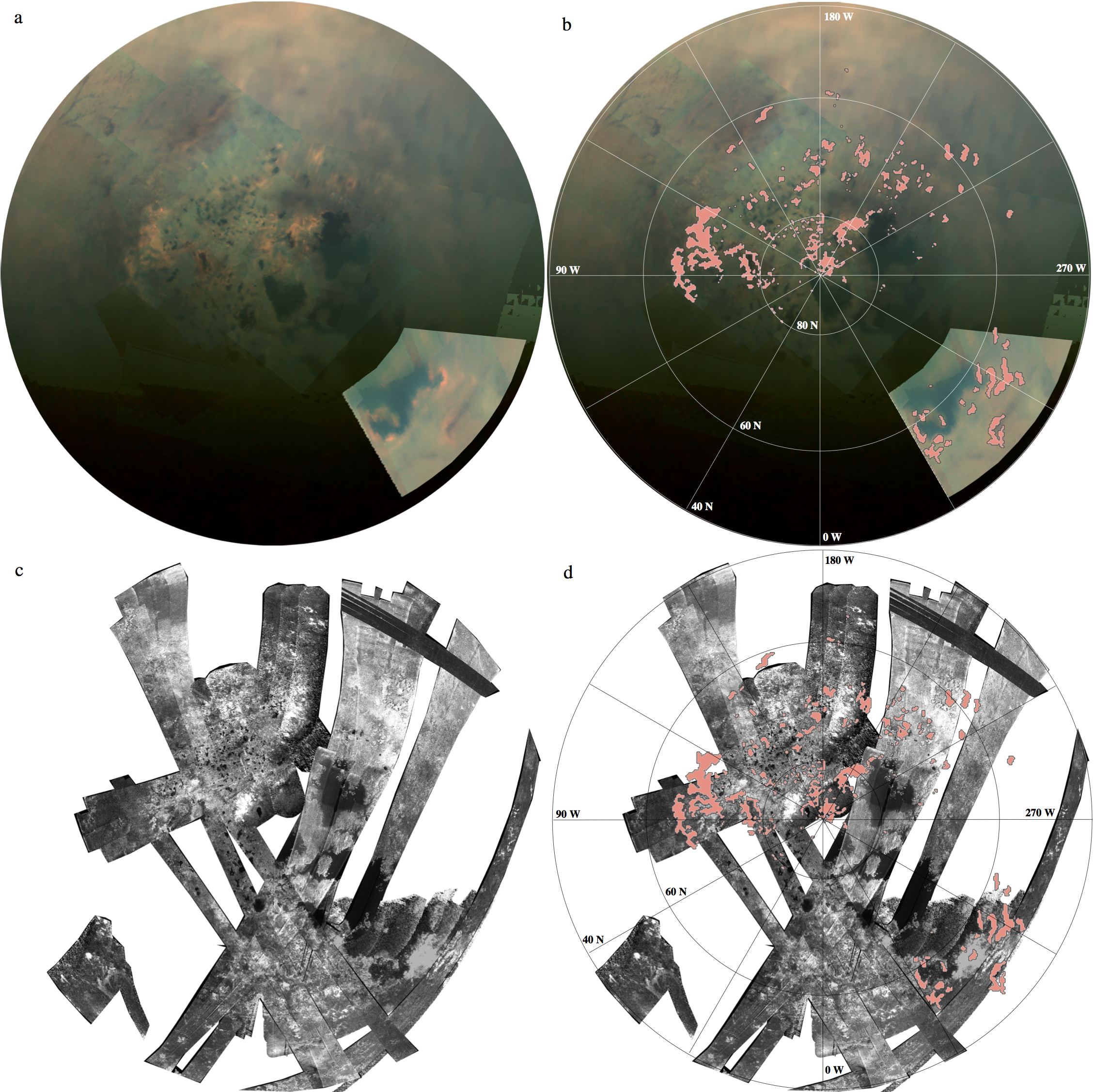}
 \caption{VIMS (T76, T93, T94, T96,T97) and RADAR (T16, T18, T19, T21, T25, T28-T30, T55, T56, T69, T83, T84, T86, and T91) coverage of Titan's north pole projected into an orthographic view from above the north pole: (a,c) without annotation and (b,d) with labeled coordinates and the polygons used to calculate the areas listed in Table \ref{evapcandtable}. While there is an observational bias towards the western half of the pole with regard to smaller lakes as seen by VIMS, it is interesting to note that there is a real discrepancy in the distribution of large seas; that is, Punga, Kraken, and Ligeia are all in the western hemisphere. On the eastern side, there are no maria, but we do see two large evaporite deposits, Woytchugga and Nakuru. The grey lines in panel b indicate the two clouds of the VIMS composite (observed in T69).}
  \label{northpole}
 \end{figure}
 
In Figure \ref{northpole}, the best VIMS coverage (panel a) of the north pole is compared to the RADAR coverage (panel c) and our mapped evaporite deposits (pink polygons of panels b and d). As Titan enters northern spring, VIMS has begun to explore the previously night-covered, near-polar surface. In addition, the dense cloud cover above the north pole that was present during northern winter has progressively vanished since equinox, allowing observations of the underlying lakes \citep{2011Icar..216...89R,2012P&SS...60...86L}. Notably, with flybys T90 and later, we were able to investigate the anti-Saturn half of the north polar region, including the extent of Punga Mare and a variety of yet unnamed lakes. \\ 

\begin{figure}
\noindent\includegraphics[width=30pc]{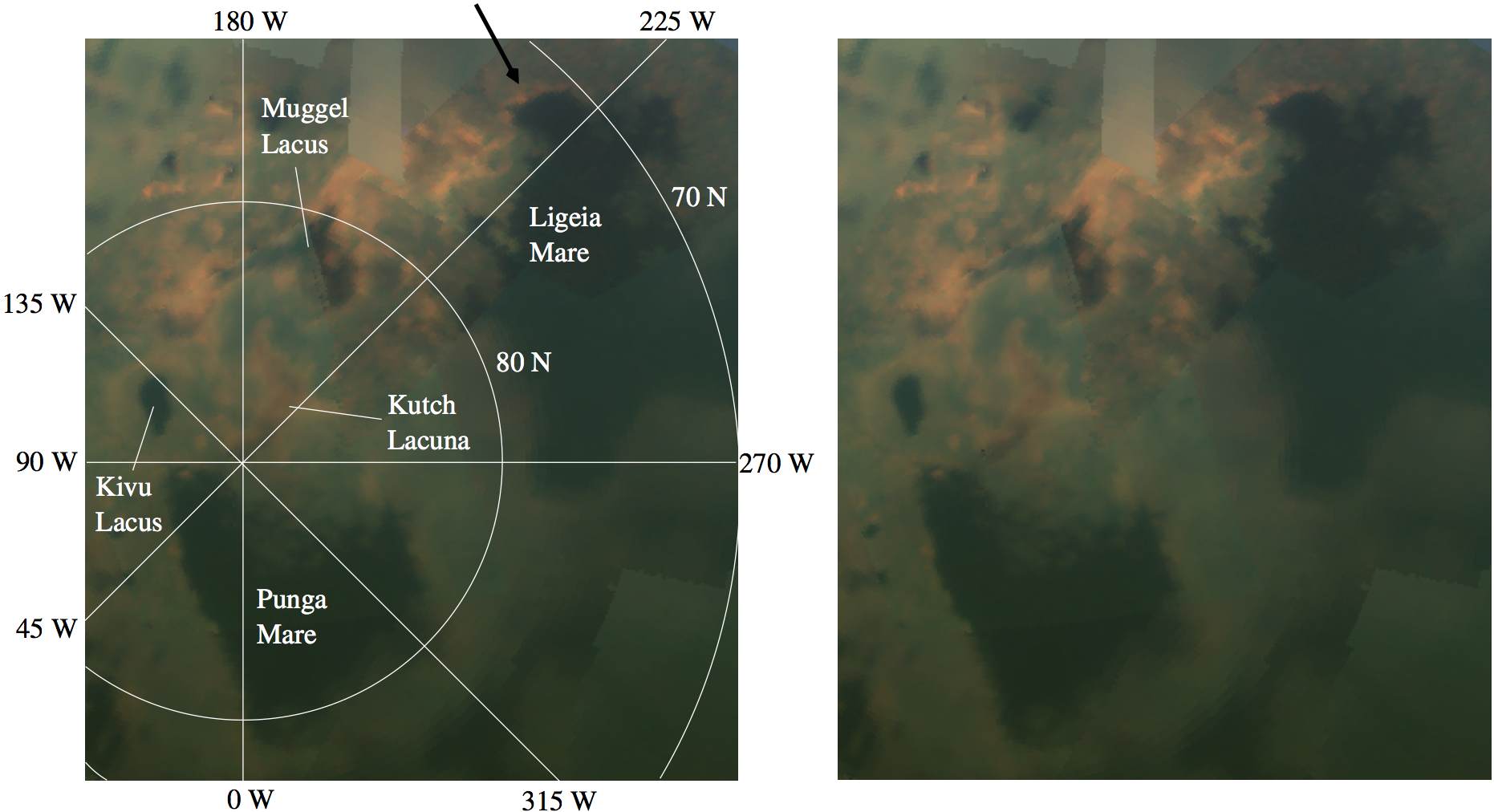}
 \caption{VIMS coverage of Titan's north pole from T93 and T94 projected into an orthographic view from above the equator with labeled coordinates (left) and unannotated (right). The ``strip" of evaporite along the shore of Ligeia that seems to connect to Muggel Lacus's evaporite is pointed to by a black arrow. It may be that Muggel Lacus, unlike Kutch Lacunae, is not an isolated system, but rather a remnant of a time when Ligeia Mare extended further north. }
 \label{northzoom}
 \end{figure}

In the latest VIMS data, we see that Punga Mare and Kivu Lacus are not the only features surrounding the exact North Pole: a distinct evaporitic feature, Kutch Lacuna, exists independently. Shown in Figure \ref{northzoom}, Kutch does not lie along the shorelines of either the lake or sea nearby. Interestingly, Punga has small evaporite deposits only visible in the high resolution shot of T93 (best sampling of $\sim$ 5 km/pixel). This is unusual, as the other three largest bodies (Kraken Mare, Ligeia Mare, and Ontario Lacus) have both small deposits and larger shoreline coverage of 5-$\mu$m-bright material. Perhaps Punga's liquid level is currently at its ``high water" mark, or the sea could have steeper shores than its counterparts. \\

\begin{figure}
 \noindent\includegraphics[width=15pc]{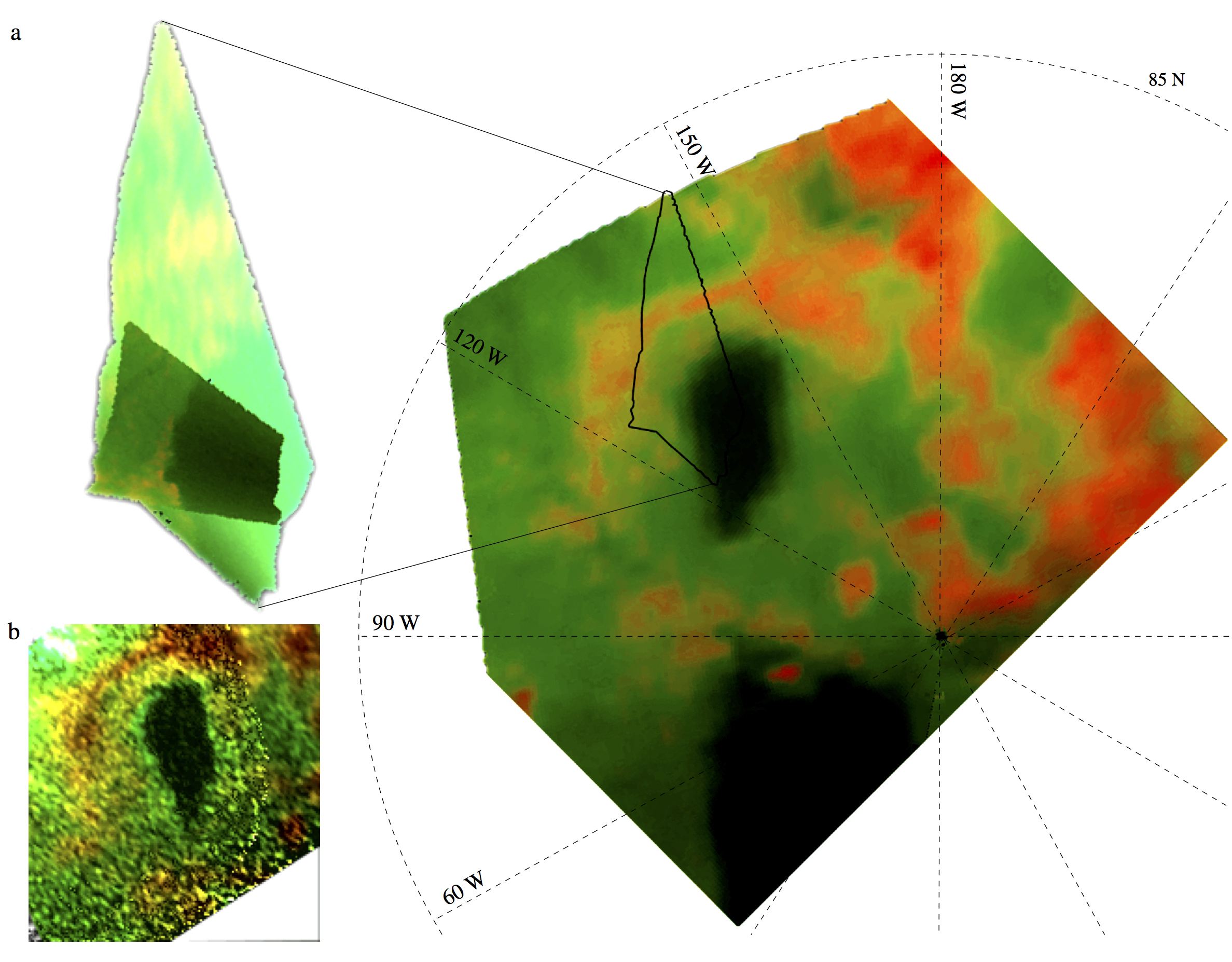}
 \caption{Orthographic projection of Kivu Lacus. (a) Annotated VIMS data from T93 is shown on the right with the outline of the high resolution image from T85, a zoom in of which is shown on the left. (b) RADAR data (HiSAR T25) combined such that the VIMS data provides hue and saturation while RADAR provides the value (HSV). We see that 5-$\mu$m-bright material (bright orange and red in this HSV) is geomorphologically consistent with evaporite. As with Titan's other lakes, evaporite is not found along the entirety of the shoreline, though we do not have the resolution around the entire shore to be certain that the evaporite is truly absent. There are some non-shoreline 5-$\mu$m-bright areas (a semicircle a few kilometers from Kivu's southern shore, seen in both T93 and T85) though no RADAR features are discernible in the low signal of available data. }
 \label{Kivu}
 \end{figure}

The southeastern shore of Kivu Lacus was seen at a fine sampling (roughly 1 km/pixel) by VIMS in the T85 flyby. In T85 (zoom in of Figure \ref{Kivu} a), 5-$\mu$m-bright material hugs the south eastern shoreline. In panel b, we use a hue-saturation-value (HSV) scheme to combine RADAR and VIMS data. The RADAR data provide the combined image's brightness (value), while VIMS data determine the color (hue) and degree of color (saturation) relative to the value. We also identify some non-shoreline 5-$\mu$m-bright material (orangey-red in the color stretch of Figure \ref{Kivu}) as deposits though no RADAR features are discernible. These deposits near the lake were observed in both T85 and T93, forming a half ring about 150 km wide a few kilometers from Kivu's southern shore. \\

North of Ligeia, a large distribution of 5-$\mu$m-bright material surrounds the uniquely shaped, VIMS dark blue lacustrine feature known as Muggel Lacus, shown in Figure \ref{northzoom}. The morphology suggests either a paleo shoreline of Ligeia Mare or a previously more extensively filled lacustrine feature. In the former case, Ligeia would extend over an area of 136,000 km$^2$, an increase of about 9\%. Muggel Lacus was covered by what we assume to be fog in T91, but returned to the morphology seen in T90 by T93 and T94. The most northern extension of Muggel Lacus (near Kivu) was covered by a cloud in T93, but is cloud free in T94. We interpret Muggel to be a shallow evaporitic basin with a very shallow lake at its bottom-- shallowness would be consistent with why the feature is difficult to see in RADAR.\\

\begin{figure}
\noindent\includegraphics[width=\textwidth]{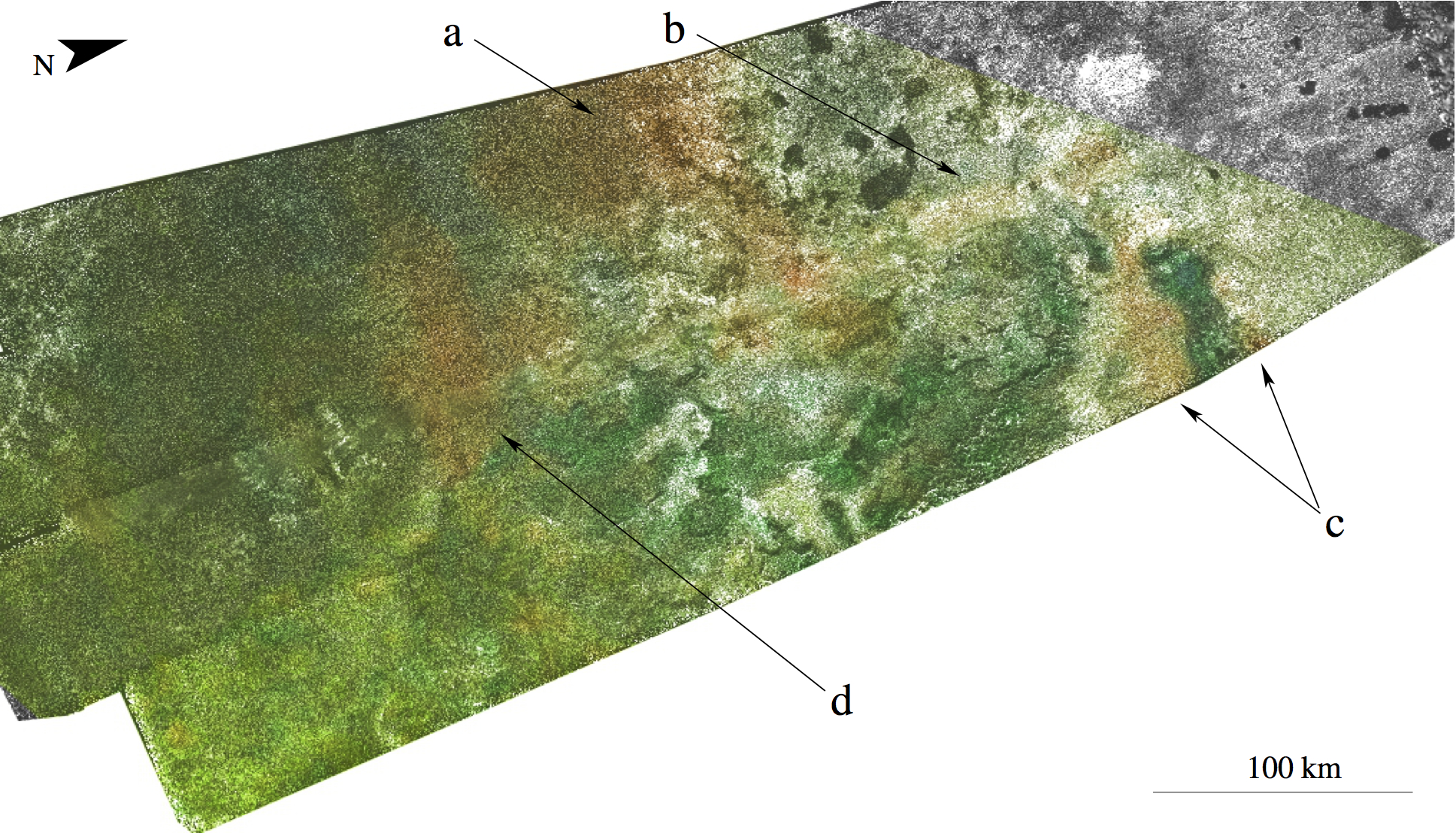}
 \caption{VIMS images of the eastern north pole evaporites from T93 overlaid atop RADAR HiSAR from T91 with an HSV color-stretch. Arrows \emph{a} and \emph{b} point to evaporite candidates, Woytchugga and Nakuru respectively, that are partially seen in a small, high resolution image from T88 (and appear 5-$\mu$m-bright there).  Woythchugga is located at 69\degree N, 110\degree W; Nakuru is at 65.5\degree N, 92.38\degree W. Evaporite surrounding MacKay Lacus is indicated by the \emph{c} arrows (77\degree N, 96\degree W). The deposit pointed to by arrow \emph{d} is seen in T90-T97 as a 5-$\mu$m-bright feature of static morphology. To date, the best resolution images of Woytchugga and Nakuru (arrows \emph{a} and \emph{d}) are from T97.}
 \label{T93T91composite}
 \end{figure}

In Figure \ref{T93T91composite}, we again combine VIMS and RADAR data and see that the expansive 5-$\mu$m-bright region indicated by arrow \emph{a}, Woytchugga Lacuna, is similar to Kutch. Neither are located in conjunction with currently filled liquid bodies. To the right of Woytchugga is another not-lake-bordering 5-$\mu$m-bright deposit, Nakuru Lacuna, indicated by arrow \emph{d}. The 5-$\mu$m-bright signature of both candidates is consistent in shape between T90 through T97-- they are thus unlikely to be patches of fog. The northeastern-most corner of Woytchugga even appears 5-$\mu$m-bright in a small hi-res shot from T88. Smaller deposits identified from T88 also agree with the deposits identified in this T93 image, \emph{b} and \emph{c}. Arrow \emph{b} points to a streak that does not seem to correspond with SAR-identifiable lacustrine features, though in VIMS data it borders the dark material typical of liquid filled lakes (blue in our color scheme) in the north pole. Arrow \emph{c}, however, shows the often observed evaporite-covered shores of MacKay Lacus. \\

\begin{figure}
\noindent\includegraphics[width=1\textwidth]{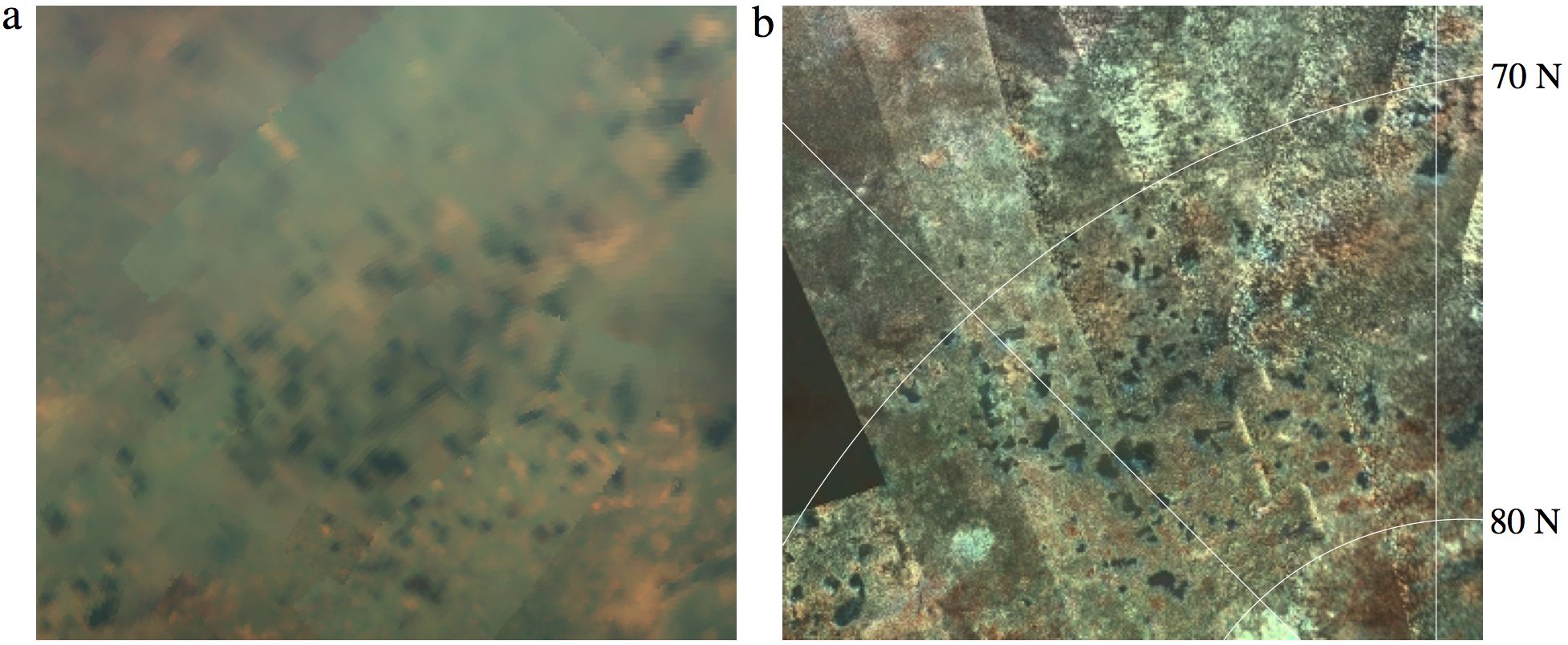}
 \caption{Titan's north polar lake district as seen by VIMS in T96 and T97 (a) and RADAR (b, colored with VIMS) in a HiSAR swath from T25. Despite VIMS' comparatively coarse resolution, the agreement between the 5-$\mu$-bright material in dry beds and around lake shores is geomorphologically consistent with evaporite. }
 \label{lakedistrict}
 \end{figure}

The region dense with small lakes identified by \cite{Hayes2008} around 80N, 140W was seen by VIMS during T96 and T97. Figure \ref{lakedistrict} shows the VIMS data on the left, and a VIMS-RADAR composite on the right. With few exceptions, the 5-$\mu$m-bright material corresponds to the shores of filled lakes or the bottoms of dried lakebeds as discerned from RADAR. Thus, evaporite is found on the largest liquid bodies on Titan (notably, the large deposits along the shores of Kraken Mare), as well as lakes at the limit of detectability for VIMS. There seems to be a higher number of evaporite deposits in this region than elsewhere in the north pole, though we caution that this statement is observationally biased until the sub-Saturn half of the north pole has better VIMS coverage.\\

While we will show in this section that much of the correlation between RADAR and VIMS data is geomorphologically consistent with evaporite formation along lakebed bottoms or liquid filled shorelines, the remaining diversity of RADAR features corresponding to VIMS 5-$\mu$m-bright material, especially in the liquid-abundant north pole region, is evidence for complex processes that are not yet understood. Woytchugga and Nakuru Lacunae, for example, are not located near current lacustrine features. Instead, the 5-$\mu$m-bright areas seem to be located along a boundary between varying intensities in the RADAR signal (perhaps similar to Arrakis Planitia, discussed below). \\

The geomorphological interpretation by \citet{Wasiak2013} of the RADAR coverage of Ligeia Mare complements the evaporite candidates that we map in the region. The deposits are located in the kinds of terrains one would expect evaporite to form. Evaporites dotting the southern shoreline of Ligeia (identified in T69), for example, are small, isolated, and coincide with bays (possibly flooded) and mottled (erosion pattern lacking) landscapes. In the area between Kraken and Ligeia, where we identify evaporite along the shore of Kraken, \citet{Wasiak2013} interpret a dome structure to establish a watershed between the three seas. The ``strip" of evaporite along the easternmost southern shore of Ligeia (viewed in T94 and pointed to by a black arrow in Figure \ref{northzoom}) is located in a lower lying region than the hummocky terrain directly east.\\

\begin{figure}
\noindent\includegraphics[width=20pc]{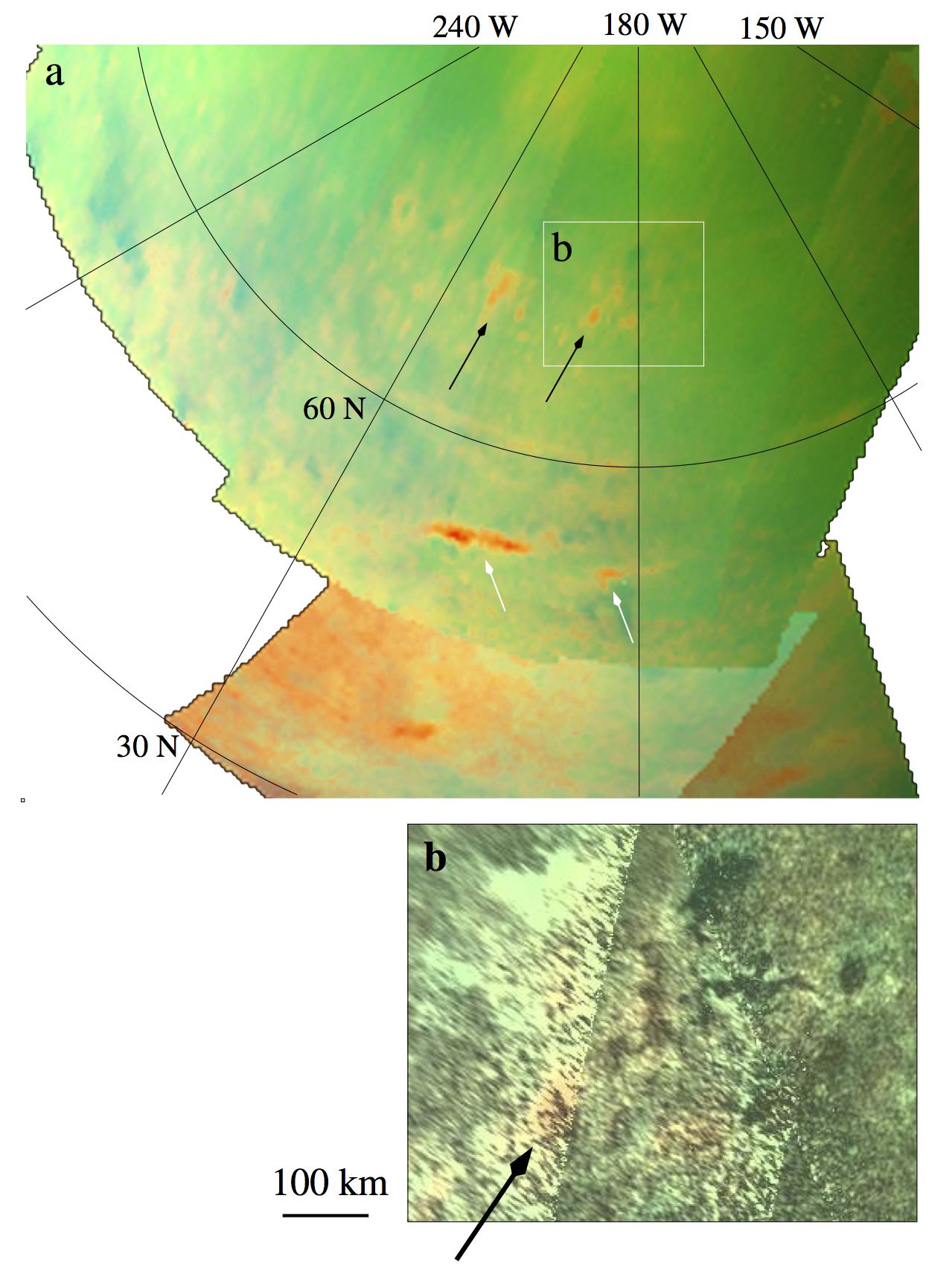}
 \caption{(a) VIMS image from T69 of the lake dotted region located north of Dilmun. The white arrows point out clouds (identified via their 2 $\mu$m  window signature). The other 5-$\mu$m-bright isolated areas are evaporite candidates. (b) HSV composite of RADAR data from T29 colored according to VIMS. Most of the HiSAR mode RADAR swath covering this region is of too poor resolution to make a comparison to VIMS. However, the section shown in (b) has high enough signal-to-noise to give, at least qualitatively, more geomorphological evidence that 5-$\mu$m-bright signatures correspond to lake features in RADAR.}
 \label{figure:Dilmun}
 \end{figure}

North of the region Dilmun, evaporite deposits continue eastward from those south of Ligeia identified by \citet{evaporite}, shown in Figure \ref{figure:Dilmun}. The two northern deposits indicated with black arrows appear in T69 as well as T90 and T97. The correlation between RADAR and VIMS data for this general area is ambiguous, as the HiSAR swath of T29 is noisy along the seams. The data do, however, provide enough qualitative context for the evaporite deposits outlined in \emph{b} to again confirm that the 5-$\mu$m-bright material coincides with a region in which lakes exist. Clouds are also active in this region: the white arrows of Figure \ref{figure:Dilmun} point to clouds that exhibit the specific 2 $\mu$m wing, 2.75 $\mu$m, and 5 $\mu$m window signatures. Whether the lake desiccation process drives the clouds or the clouds drive the precipitation is beyond our present ability to determine.\\

\begin{figure}
\noindent\includegraphics[width=20pc]{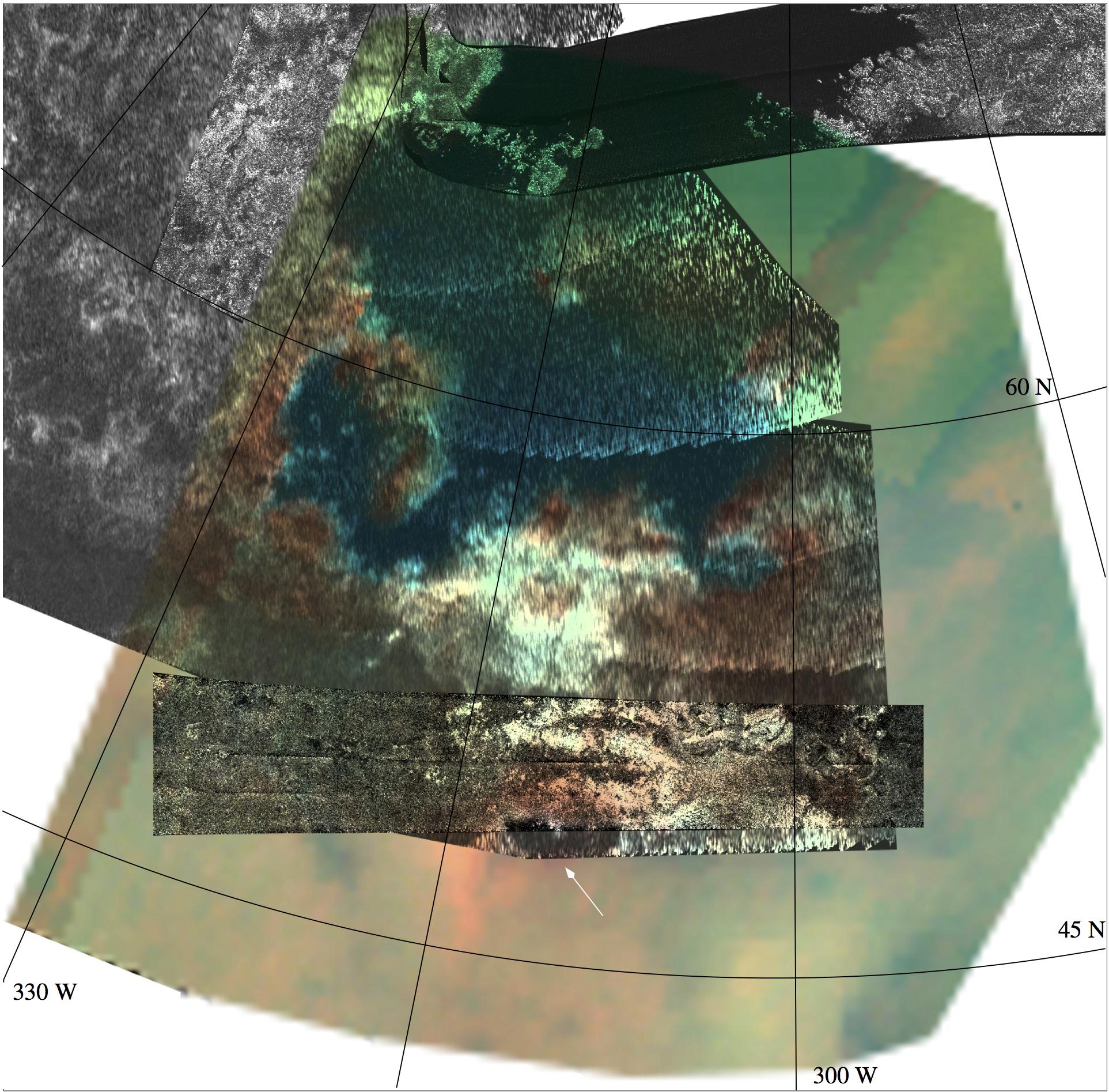}
 \caption{(a) Best resolution VIMS coverage of the southern shoreline of Kraken Mare from T76 overlaid with SAR and HiSAR data from T84 in HSV space. The evaporite deposits (explicitly outlined in Figure \ref{northpole}b) correlate well with the sea's shoreline, as well as with the boundary of the RADAR-dark feature to the south and a depression-like area to the east. The white arrow points to Hammar Lacus; the RADAR-dark material indicates a smooth surface coinciding with dark VIMS material typical of liquid-filled lakes.}
 \label{leKraken}
 \end{figure}

Kraken Mare, Titan's largest body of liquid, extends further south than any other sea. In fact, the highest resolution VIMS imaging of the liquid body covers its southernmost shore. In Figure \ref{leKraken} we again use HSV to compare VIMS and RADAR data. The difference in quality (mostly spatial resolution) between the middle RADAR swaths and those at the top and bottom is the difference between HiSAR and SAR modes, respectively. As with Ontario Lacus \citep{BarnesOntario}, evaporite is not detected along the entirety of the shoreline seen at VIMS resolution. If the RADAR-bright material of the southwest is a mountain chain, then the steep shoreline could explain the apparent absence of evaporite along the middle of the southern coast. This would be a scenario similar to the northwest shore of Ontario, where mountain chains have been identified \citep{2010GeoRL..3705202W,Hayes2011,Ontario.dries.up,Ontario.Etosha}. On the northwest shoreline above 60\degree N, there is neither extensive evaporite coverage (only an isolated deposit viewed in T69) nor RADAR-bright material to explain the absence of 5-$\mu$m-bright material, similar to the other mare shorelines. Unless there are small deposits beyond presently available VIMS resolution, some other process, like steeper shorelines or frequent flooding for example, could be responsible for preventing large scale evaporite formation there.\\ 

South of Kraken, there are several evaporite deposits not located along the sea's shoreline. In the middle of the shoreline mountains, what may be a valley is seen to be 5-$\mu$m-bright. To the east, the 5-$\mu$m-bright material that appears to be separate from the very bright deposits on Kraken's southeastern-most shore extends from 58\degree N down to 50\degree N. At the lower latitudes, where SAR data is available, this material is coincident with some discernible RADAR features. Hammar Lacus, indicated by the white arrow of Figure \ref{leKraken} is about 11,000 km$^2$ in area with 5-$\mu$m-bright material delineating all but its eastern shores. The lake itself is RADAR-dark and the evaporite is RADAR-neutral to the west, as expected. To the northeast of Hammar Lacus, however, evaporite extends into the increasingly RADAR-bright area, probably indicative of mountains or very rough terrain. Mountains typically appear dark blue in our VIMS color scheme \citep{BarnesMountains}. Thus, if the RADAR-bright features are mountains, then one mechanism for them to appear 5-$\mu$m-bright would be if some erosional process was exhuming evaporite. If Kraken were to cover all these non-shoreline deposits (difficult if the RADAR-bright material is indeed mountainous), the sea would need to increase in area by about 10\%. \\

\subsection{Midlatitudes and Tropics}
\begin{figure}
 \noindent\includegraphics[width=0.7\textwidth]{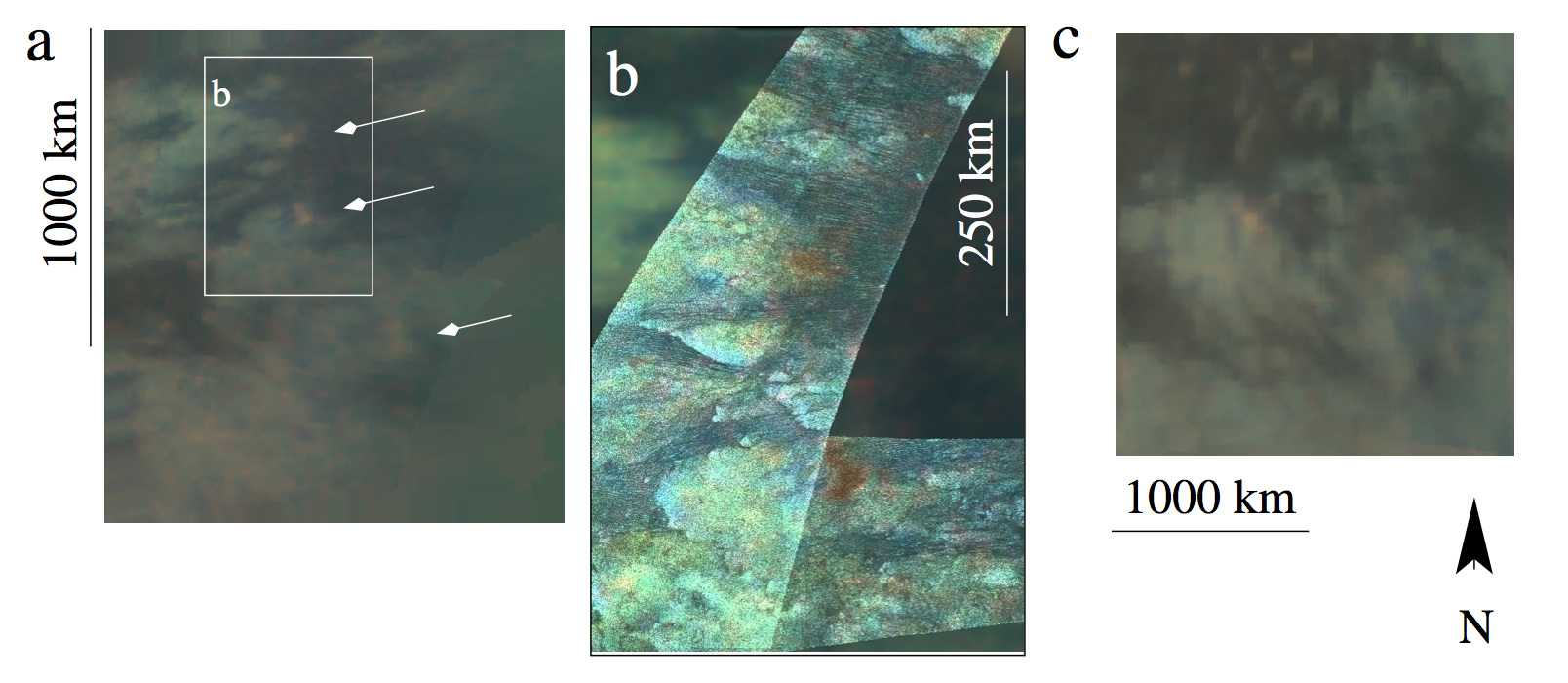}
\caption{Best VIMS images of the smaller tropical evaporite deposits: (a) west Belet north (top; 1.5\degree S, 284.9 W), south (middle; 8.2\degree S, 283.9\degree W), and southwest (bottom; 18.2\degree S, 276.6\degree W) from T61; (b) RADAR images from T8 and T21 for west Belet north and south colored according to the VIMS data; and (c) north (left) and east (right) Yalaing  from T67. All of the evaporite deposits seem to be located at the border between VIMS dark brown and VIMS equatorial bright spectral units. Where RADAR coverage exists, these VIMS boundaries coincide with RADAR dark and bright boundaries.}
 \label{figure:midlats}
 \end{figure}

In the non-polar regions, despite the present-day lack of filled lakes or seas, we identify several evaporite candidates. West of Belet, there are three distinct deposits seen in VIMS data from T61, indicated by white arrows in Figure \ref{figure:midlats}a: from top to bottom, West Belet North, West Belet South, and Southwest Belet. Each deposit seems to be located on the border between (non-Xanadu) equatorial bright and dark brown spectral units. In the equatorial band, the dark brown spectral unit, ascribed to water-ice-poor material and associated with organic material (atmospheric aerosol-like particles), corresponds to dunes \citep{2007P&SS...55.2025S,2007mapping,StephaneSinlap,Rodriguez2014}. When the RADAR image is colored according to the VIMS data (Figure \ref{figure:midlats}b), the spectroscopic boundaries correspond to those in the RADAR data and West Belet North and South are located along such a boundary. It could be that lacustrine or fluvial features are present in the region, but beyond the resolution limit of SAR, VIMS, or ISS (as is the case for the Huygens Landing Site, where probe data reveal such features while Cassini instruments do not). On the other hand, the lack of fluvial features may be indicative of distinct processes responsible for creating or exhuming evaporite along the border of the two spectral units (perhaps dune induced convecting clouds, for example \citep{BarthRidgeClouds}).\\

Unique among 5-$\mu$m-bright deposits not found at the poles, the evaporite candidate located on the north end of the land mass Yalaing Terra has been observed by VIMS several times. This evaporite candidate does not always appear 5-$\mu$m-bright: for over a year, the surface experienced brightening in the shorter wavelengths that correspond to the Equatorial Bright spectral unit, then reverted to its original spectrum, which is shown Figure \ref{figure:midlats}c. Based on the long time scale of reversion and the relative reflectivity behavior (between the brightened area and nearby unchanged dark and bright areas), this phenomenon was suggested by  \citet{PrecipSurfaceBrightening} to be due to the rainfall from the 2010 September cloudburst. Similar brightening/darkening was seen at locations of similar latitude: Heptet Regio, Concordia Regio, and Adiri. \citet{2011Sci...331.1414T} also attribute observed surface darkenings to a change in surface wetness. If rainfall is the cause for surface brightening, the evaporite may have dissolved, then precipitated out again as the liquid evaporated. Alternatively, there may have been just enough liquid accumulation to effectively mask the brightness at 5 $\mu$m, without actual dissolution of the solute. In either case, this observation is evidence against short-term evaporite volatility, for the deposit reappears in the exact location where it was previously observed. Such behavior would be unusual for a vapor. As of yet there are no RADAR data to provide further evidence as to whether this 5-$\mu$m-bright signature coincides with lacustrine or fluvial morphologies or is located along a RADAR bright/dark border in the same fashion as the other not-polar evaporite candidates.  \\

 \begin{figure}
 \noindent\includegraphics[width=0.7\textwidth]{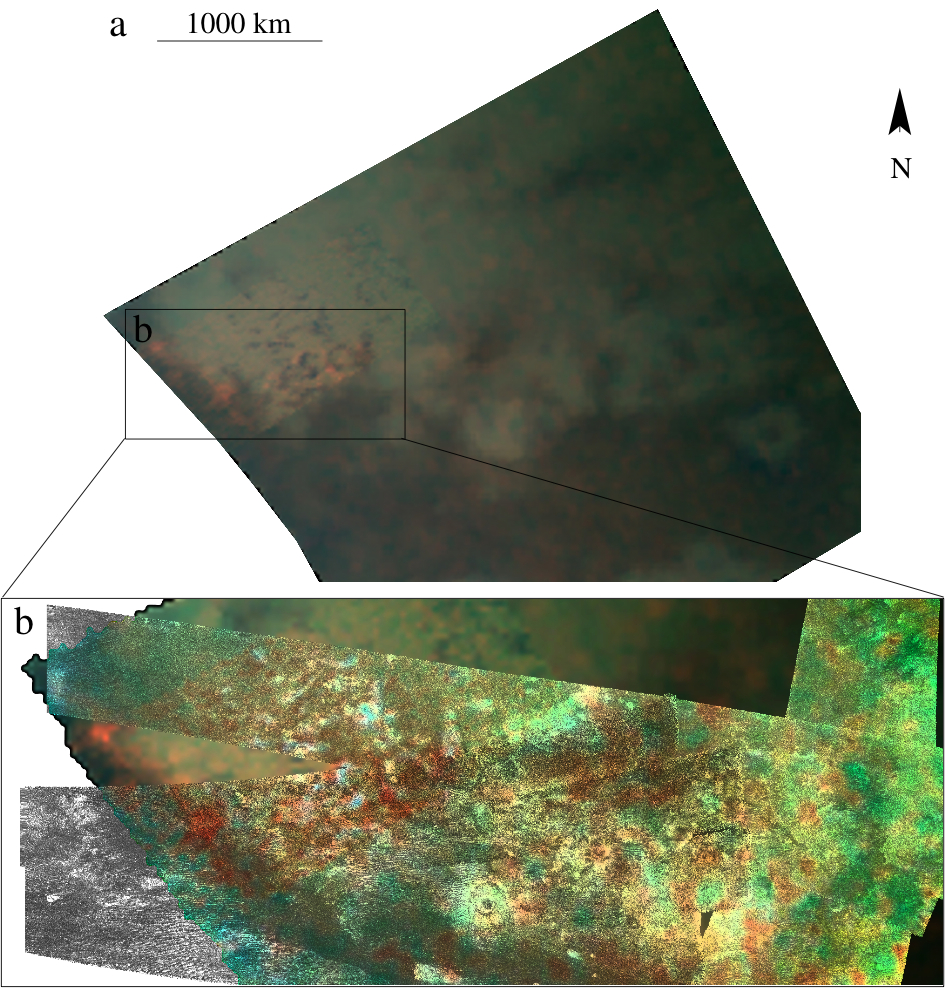}
\caption{(a) VIMS image from T5 for west and east Fensal (centered at 1\degree5 N, 52\degree W and 17.4 N, 39.5\degree W respectively). West Fensal is enclosed in the black box while east Fensal is indicated by the white arrow. (b) RADAR images from T5, T28, and T29 overlaid with VIMS in HSV space. While the evaporite candidates are easily discernible in the VIMS data alone, the high photon shot noise makes comparison to RADAR difficult. However, the VIMS dark blue material that the evaporite deposits border are seen to correlate with RADAR bright areas identified as peaks in SARtopo.}
 \label{figure:fensal}
 \end{figure}

There is a particularly bright outcropping of 5-$\mu$m-bright material north of the western part of the region known as Fensal, best seen by VIMS in T5 and encapsulated by a black box in Figure \ref{figure:fensal}a. To the furthest west of this box, two deposits are located on the border between equatorial bright and dark brown material, similar to what we see in West Belet (Figure \ref{figure:midlats}a). East of these features, there are a number of 5-$\mu$m-bright patches nestled amongst VIMS dark blue spots. The VIMS dark blue unit, unlike the dark brown unit, is suspected to be water-ice rich \citep{Rodriguez.landingsite,Rodriguez2014,2007P&SS...55.2025S,2007mapping,StephaneSinlap}. Comparison between the VIMS and RADAR data in panel b is hindered by the high level of photon shot noise; that is, the low signal of this flyby is highly affected by random photon fluctuations. While the composite is not very useful for distinguishing the RADAR features corresponding to 5-$\mu$m-bright deposits, we can discern that the dark blue VIMS unit coincides with RADAR bright material. These dark blue spots have a lacustrine-like morphology; liquid lakes could appear RADAR bright if there were enough waves to roughen the surface. However, SARtopo inferred relative altimetry \citep{2009Icar..202..584S} indicates that peaks in this region correspond to the RADAR-bright and VIMS dark blue spots. While such altimetric evidence does not rule out the possibility of lakes (there are mountain lakes on Earth, such as tarns and cirques, that are formed in the basins excavated by glaciers and filled with rainwater or snow melt), the prevalence of the VIMS dark blue unit coinciding with mountains in the equatorial region as well as the putative absence of equatorial lakes makes such an explanation less likely. Thus, we prefer an interpretation in which the peaks are mountainous and evaporite candidates are either exhumed by crustal activity or is the remnant of liquid that pooled in slight depressions at the base of the mountains, where the mass of the peaks could deform the surface crust enough to create very localized basins.\\

Further east we see a larger but fainter 5-$\mu$m-bright region (pointed to by the white arrow in Figure \ref{figure:fensal}) once again bordering equatorial bright and dark brown VIMS spectral units. The spatial resolution of this part of the VIMS composite image is not as good as that of the smaller west Fensal deposits discussed above. SARtopo indicates that this region is of lower altitude than the RADAR bright spot bordering on the east-- the topography expected for an evaporitic deposit as liquid could pool in the depression. If there was once enough liquid to form this evaporite candidate along the edges of this basin, perhaps there was also enough to wet the surface at the base of the VIMS dark blue mountains nearby.\\

\subsection{Tui and Hotei Regiones}
\label{tuihote}
Recent work by \citet{2013JGRE..118..416L} reestablishes the evidence for possible cryovolcanic features around Hotei from the most recent VIMS and RADAR data. While we maintain the hypothesis of \citet{Moore.Tui.Hotei.Lakes} that the basins Tui Regio and Hotei Regio are 5-$\mu$m-bright because they are fossil seabeds that were once filled with enough liquid from which evaporite could form, we note that cryovolcanic activity and a fossil seabed are not mutually exclusive. Isla Incahuasi, for example, is an ancient volcano in the middle of the Salar de Uyuni, Earth's largest salt flat. However, since the 5-$\mu$m-bright spectral unit is water-ice-poor, the \emph{formation} of the 5-$\mu$m-bright material is not consistent with cryovolcanism. \\

As shown in Figure \ref{figure:fractions}, Tui and Hotei are the largest single outcroppings of 5-$\mu$m-bright material on the surface of Titan, comprising 30\% and 21\% of the total area of mapped 5-$\mu$m-bright material respectively. (For VIMS observations of the regiones, we refer to Figures 1 and 2 of \citet{westTui} for Tui and Figure 1 of \citet{2009Icar..204..610S} for Hotei.) We estimated the volume that the basins could hold by a simple approximation similar to \citet{LorenzSurfaceInventory} and \citet{LorenzThroat}: we assumed that the depth/area ratio is constant for large bodies and scaled this ratio to the recently measured depth value for Ligeia Mare \citep{LigeiaDepth}. From this, the total volume in the seas was found to be $\approx$ 42,000 km$^{3}$, a value within the range of those found independently by \citet{LorenzThroat} and \citet{2014LPI....45.2341H} ($\approx$ 32,000 km$^{3}$ and  $\approx$ 70,000 km$^{3}$, respectively). Our total volume for Tui (ratio depth of 370 m) and Hotei (200 m) was 57,000 km$^{3}$. To first order, there is enough liquid currently in the north pole to fill Tui and Hotei.

\subsection{South Pole}
\begin{figure}
\noindent\includegraphics[width=\textwidth]{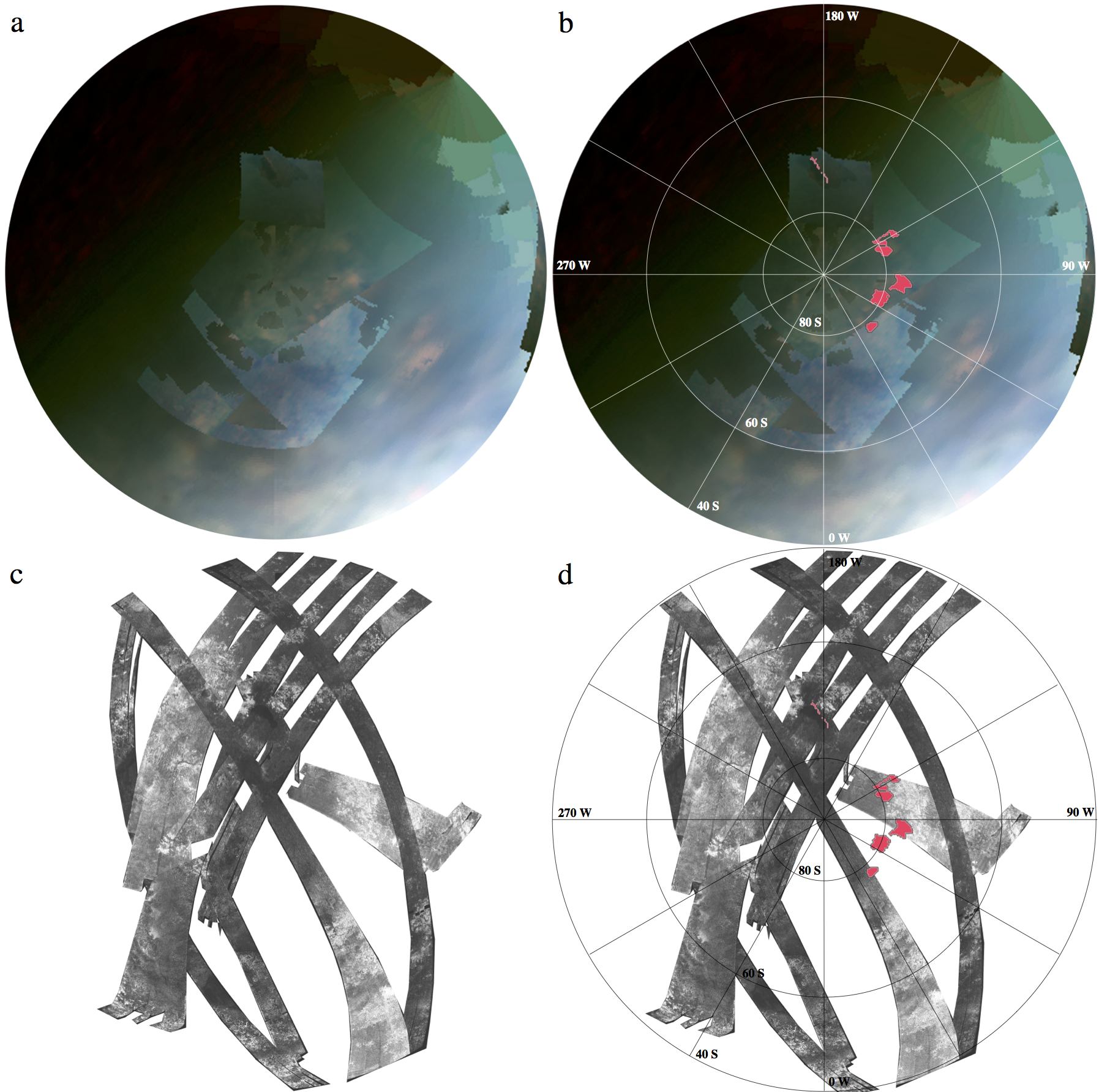}
\caption{Orthographic projections of Titan's South Pole as seen by VIMS in panel a (T20, T23, T24, T28, and T51), by RADAR in panel c (T36, T39, T49, T55, T56, T57, T58, T59, T65, and T71), and annotated with polygon outlines of evaporite candidates in panels b and d.  Clouds have been removed from this VIMS composite to increase surface visibility. Fewer lacustrine features are present in this region in comparison to the north pole, as are evaporite candidates. Ontario Lacus, seen along the 180\degree W line, is the region's only known long-lived liquid body with evaporite. Deposits on the edge of Arrakis Planitia, however, are seen by VIMS after an ISS identified surface wetting event \citep{2009GeoRL..3602204T}. (There is are no VIMS observations of the area before those of ISS.) The other evaporite candidates, though they lack RADAR data with which to correlate, are observed in multiple VIMS flybys, in the midst of frequent cloud activity.}
\label{southpole}
\end{figure}

Cassini's RADAR has revealed Titan's south pole to have a geomorphology as complex as the north: hummocky terrain thought to be topographic highs \citep{Lopes2010}, fluvial networks \citep{Burr2013742}, and lacustrine features both persistent and ephemeral \citep{Hayes2011}. Unlike the north pole, however, where lacustrine and evaporitic features are widely distributed and vary greatly in size, the southern polar region is dotted with only a handful of evaporite candidates large enough for VIMS to spectrally identify. Shown in Figure \ref{southpole}, these deposits are found along the shorelines of Ontario Lacus, isolated in the eastern half of the region, and on the edges of Arrakis Planitia. \\

The 5-$\mu$m-bright material along the eastern shore of Ontario Lacus, the largest and most long-lived liquid filled body in the region \citep{T38.ethane, 2009GeoRL..3602204T}, has been well documented \citep{BarnesOntario,2010GeoRL..3705202W,Ontario.dries.up,Ontario.Etosha,Ontario.nochange}). The general geomorphological interpretation is that this area is an evaporite covered \citep{BarnesOntario} alluvial plane \citep{2010GeoRL..3705202W,Ontario.dries.up,2010GeoRL..3705202W}, though \citet{Ontario.Etosha} also propose lunette dunes. We estimate that evaporite covers an area equivalent to 17\% of Ontario's liquid filled portion and would require an increase in liquid volume of about 30\% to cover it (using the same method described in Section \ref{tuihote}). \\ 

\begin{figure}
\noindent\includegraphics[width=15pc]{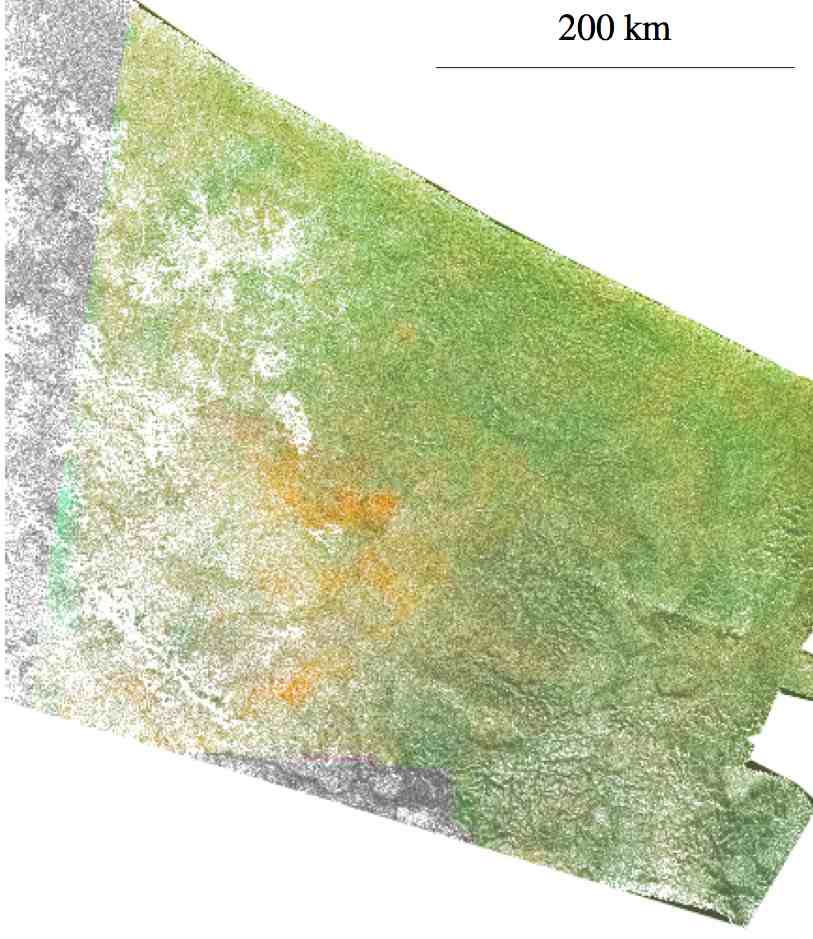}
\caption{RADAR image of Arrakis Planitia (78.1\degree S, 110\degree W) from T49 orthographically projected and colored according to VIMS data from T51. The 5-$\mu$m-bright material is coincident with the edges of the low plain whose albedo was observed to darken after presumed rainfall from earlier cloud coverage. ISS detected a return to the original albedo in T50 \citep{2011Sci...331.1414T}. VIMS observes the 5-$\mu$m-bright signature in several flybys, but only after the darkening period.}
\label{arrakis}
\end{figure}

A darkening of Arrakis Planitia was observed from T0 (2004 July 3) to Rev009 (2005 June 6) by ISS coincident with cloud coverage in Ta (2004 October 6) \citep{2009GeoRL..3602204T}. The area was then seen to return to the lighter albedo observed in 2004 in ISS data procured in T50 (2009 February 6) \citep{2011Sci...331.1414T}. This surface change was interpreted by \citet{2009GeoRL..3602204T} and \citet{2011Sci...331.1414T} as precipitation driven wetting and subsequent drying by infiltration or evaporation. VIMS observed Arrakis Planitia in T20 (2006 October 25), T22 (2006 December 28), T23 (2007 January 13), T24 (2007 January 29), and T51 (2009 March 27), revealing some 5-$\mu$m-bright signatures in the same area. In T20, the signature is small and faint. In T22, the area is covered by low altitude clouds, which seem to move or dissipate by T23. The 5-$\mu$m-bright feature looks the same in T23 and T24, but by T51 extends over a somewhat larger area. We interpret these observations as complementary to the processes observed by ISS--the increase in 5-$\mu$m-bright signature could be due to surface liquid (either still present from Ta or in connection to the cloud activity of T22) evaporating and leaving the 5-$\mu$m-bright material behind. This would be similar to the scenarios explored by \citet{PrecipSurfaceBrightening}, where the evaporite deposit north of Yalaing Terra disappears after a rainfall event and reappears when (presumably) the surface liquid has evaporated again. Figure \ref{arrakis} shows the  correlation between RADAR coverage from T49 and the VIMS data from T51. The evaporitic material lies along the border of the low plain. (The T24 coverage of Arrakis Planitia is shown in the VIMS composite of Figure \ref{southpole}.) \\

The other two candidates in the south pole do not coincide with any RADAR data, but are seen to be 5-$\mu$m-bright in more than one flyby. We note that the locations of our evaporite candidates, except for that around Ontario Lacus, do not coincide with the pole's large basins \citep{2012DPS....4420108S,LorenzTopo} proposed by \citet{2013LPI....44.1764W} as large enough to be seas if they were filled. If the surface liquid transport were mostly seasonal, we would expect to see these basins filled to some extent during Cassini's observations of southern summer. The absence of such filling in addition to the lack of evaporite would point to some longer than seasonal process of liquid exchange, if any ever occurred. The small lakes of the south pole \citep{Hayes2011} could have a more seasonal cycle, though we anticipate that such a relatively short period of fill would not be enough to accumulate enough solute for evaporite formation. \\

\subsection{Topography}
\label{toposection}
\begin{figure}
\begin{tabular}{c}
\includegraphics[width=0.6\textwidth]{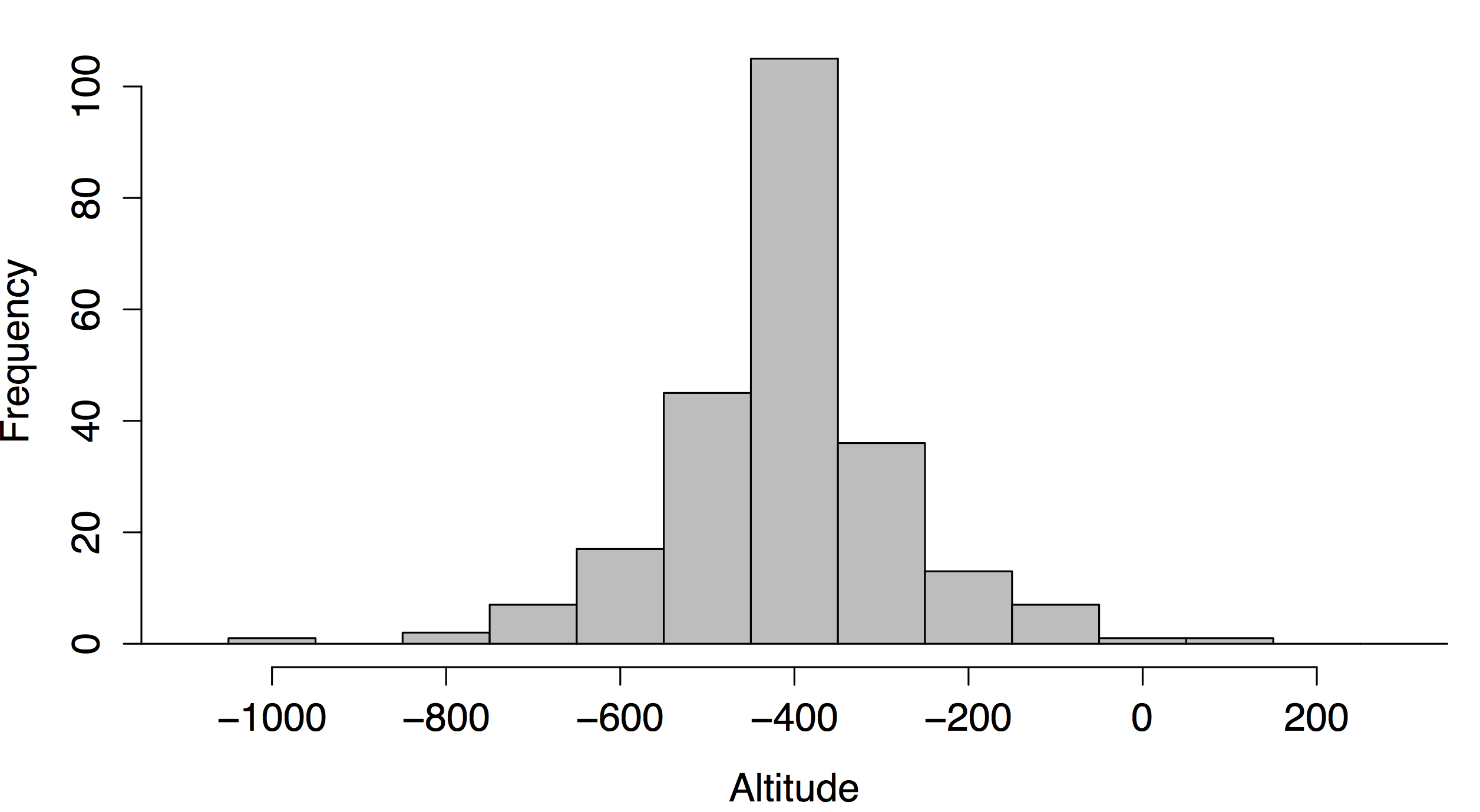}\\
\includegraphics[width=0.6\textwidth]{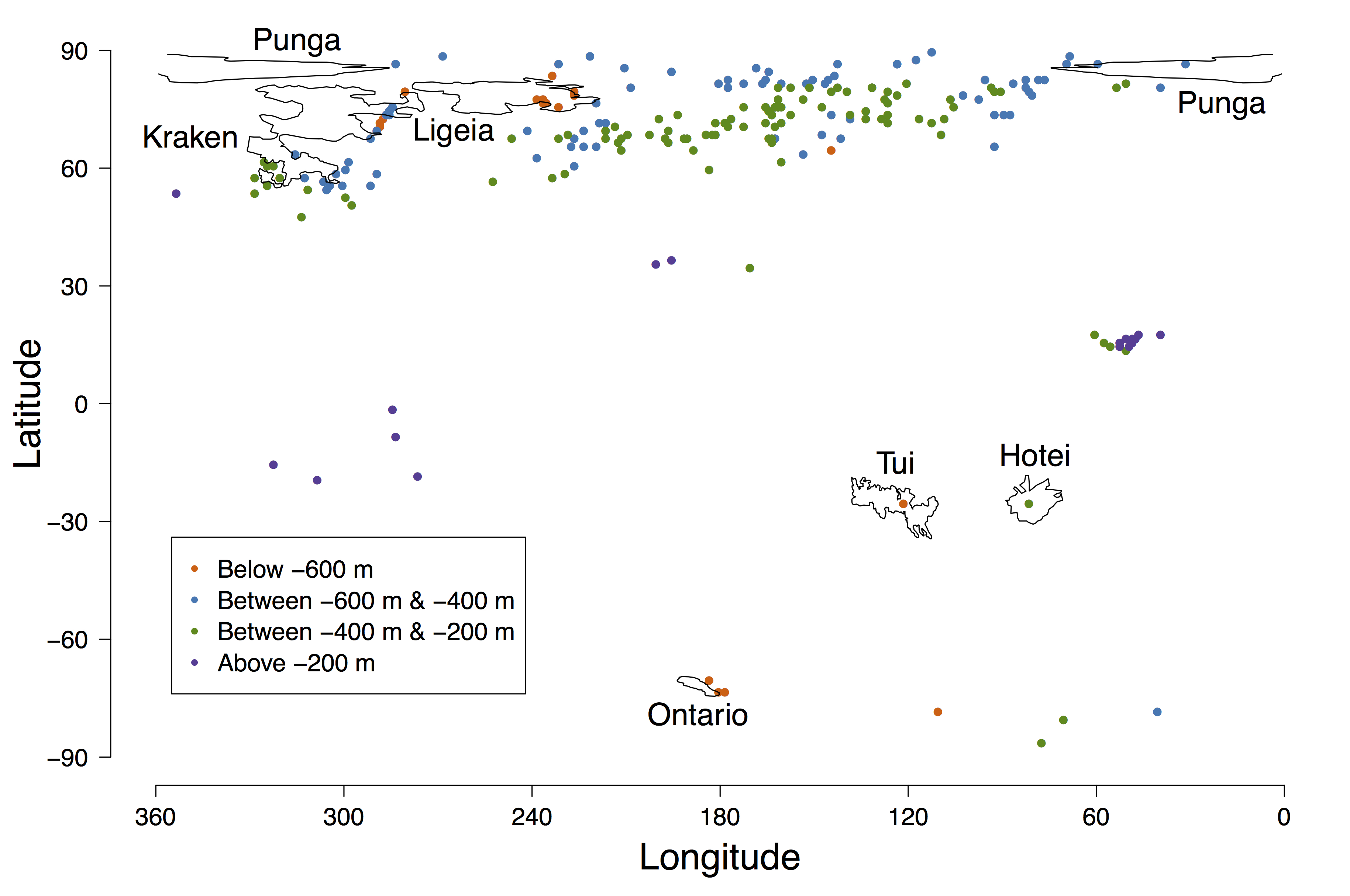}
\end{tabular}
\caption{Altitudes of the evaporite candidates from the topographic map of \citet{LorenzTopo} relative to the degree 3 geoid of \citet{2012Sci...337..457I} shown in (a) a histogram and (b) in cylindrical projection with contours of major bodies of liquid drawn from VIMS data. The average altitude is about -400m with $\sigma$=140m. }
\label{topofig}
\end{figure}

To date, the available topographic data for Titan cover only 11\% \citep{LorenzTopo} of the surface area. Recent work by \citet{LorenzTopo} interpolates a global topographic map for Titan based on these data from RADAR SAR, altimetry, and SARtopo \citep{2009Icar..202..584S}. To investigate the relative topography of our evaporite candidates, we transform the data of \citet{LorenzTopo} into altitudes relative to the degree-three geoid of \citet{2012Sci...337..457I} by using the ellipsoid defined by $(a,b,c)=2575+(230,-68,-171)/1000$. The results are shown in Figure \ref{topofig}: panel a shows the frequency of altitudes while panel b shows the global distribution of evaporite candidates binned by altitude.\\

The Gaussian shape of the histogram shows that the 5-$\mu$m-bright material is not always at the overall lowest points on Titan. This is perhaps unsurprising in the case of evaporite formation. Liquid pools in \emph{local} topographic lows, not necessarily global lows. After all, the largest salt basin on Earth today, Salar de Uyuni, was once a filled lake at 3,600 m above sea-level in the Andes Mountains. Additionally, salt deposits on Earth can be buried and later exhumed by tectonic activity. Comparing our evaporite candidate locations to the immediate vicinity indicates that the deposits are locally low-lying, at least at the resolution of the interpolated map. Further investigation and quantification of our proposed evaporite covered topography will therefore have to wait for a more conclusive topographic data set.\\

\section{Discussion}
\label{discussion}

Based on the correlation between 5-$\mu$m-bright material and lakebeds shown in this work and by \citet{evaporite}, we explore the implications of liquid having been present at some time in Titan's history at the locations now covered with 5-$\mu$m-bright material (with the caveat that more of this material may be present but buried under a thick enough layer to mask its spectral signature.) In this section, we begin with a discussion of explicitly evaporation-formed deposits, as the geomorphological evidence for 5-$\mu$m-bright material on the shores of lakes in the north pole strongly suggests such a process has taken place on the surface (regardless of whether all our 5-$\mu$m-bright deposits are formed in this exact manner or not). We then discuss the different scenarios implied by the absence of evaporite deposits in the south pole and the presence of such material in the equatorial basins Tui and Hotei.  We conclude with a comment on the seasonality of these processes. \\

\subsection{Evaporite Formation}

The drying-up of a liquid body does not always involve the creation of evaporite: evaporite only forms when a solution becomes saturated and the dominant mechanism for liquid removal is evaporation. This is more likely to happen in closed systems where solute can't be washed downstream or the solution can't percolate through a porous regolith. In our discussion we describe such systems as long-lived, i.e. having been closed systems of surface liquid long enough to enable saturation of the solution.  Short-lived surface liquids, such as those left after a rainfall, may come into contact with a large amount of dissolvable material as they flow across the surface. However, maintaining saturation and evaporation-dominant removal is probably more difficult, especially if the flow is fairly fast or in contact with other streams of liquid. We note that ``short-lived" is an ambiguous term, referring to an amount of time that is simply less than that required for saturation to take place-- the latter is a function of the chemical nature of both solute and solvent, as well as the terrain over which these surface processes are happening. For example, \citet{Malaska2014} suggest that an ethane-rich Ontario Lacus should have reached saturation with benzene falling from the atmosphere within 3-20 million years.  \\

Due to the yet-unknown exact chemical composition of Titan's evaporites, we cannot definitively exclude the possibility that the deposits are themselves volatile. As Cassini has yet to capture even one full Titan year and data for particular regions are not continuous, it is difficult to observationally determine the rates of such processes. We expect, however, that they would have to operate on geological timescales based on the relative stability observed of the evaporite candidates: the shoreline deposits around Ontario Lacus are seen in T38 and T51 (a little over one terrestrial year); Kraken Mare's shoreline deposits are observable from T61-T82 (two years); and Tui Regio is 5-$\mu$m-bright from Ta-T51 (four and a half years). The evaporite deposit at the north end of Yalaing Terra did disappear for some time following a rain induced surface wetting event, but then returned to its original, 5-$\mu$m-bright spectral signature. This would seem to imply that if the 5-$\mu$m-bright material is volatile, it either recondenses onto the surface in the exact same location (within 3-11 months after the first observation of the evaporite's disappearance) or is only volatile on longer time scales. There is no evidence from the Cassini data that the 5-$\mu$m-bright material is volatile on a less than seasonal time scale, though we cannot rule out longer time scales without a better definition of the material's chemical makeup.\\ 

\subsection{South Polar Basins}
\citet{2009NatGe...2..851A} argue that the periodicity of which pole receives the greater summer insolation (according to the behavior of Saturn's orbit) is responsible for the migration of liquid between the north and south poles (a Milankovich cycle). Thus orbital influence driving the evaporation-to-precipitation ratio assumes that the south will at some time look like the north does today: the region where the majority of the surface liquid is concentrated. Why then, if the south at one time housed enough liquid to make seas from the dry basins observed today \citep{2013LPI....44.1764W,LorenzTopo} are there no evaporite deposits similar in extent to the north pole? We offer two categories of answers to this question: either (1) evaporite formed but is no longer visible or (2) conditions were never suitable for evaporite formation.\\

The southern basins could be closed systems; they are probably the lowest points \citep{LorenzTopo} in the region, so liquid would tend to pool there. Thus, evaporite formed at the south pole would have either been covered up since the liquid moved north or was itself volatile. If the evaporite is simply covered by settling atmospheric aerosols, a 0.15 -1.5 mm thick layer could build up over the 45 kyr period predicted for a Milankovich cycle. This is an upper limit-- the time required to completely empty the southern seas is not taken into account. As VIMS only probes a few microns into the surface, this could be enough to mask an evaporite layer. It could also be that the 5-$\mu$m-bright material is ephemeral: eroded away by weather, volatile enough to sublimate, or chemically altered while exposed on the surface. All of these processes could occur on Milankovich cycle timescales to remove 5-$\mu$m-bright material.\\

And yet, another explanation for the lack of evaporites is that they never formed. Perhaps the basins were not filled long enough for the liquid to reach saturation. Or, some other mechanism may have been more dominant in removing liquid from the system-- a porous regolith or connection to a subsurface reservoir could empty the basins without any evaporite deposition. These scenarios are inconsistent with Milankovich cycles, as there is no evidence of a subsurface connection to allow the liquid to get back to the north pole \citep{SotinEquipotential}. \\

Evaporite also wouldn't form if the basins have never been filled (clearly not in agreement with a Milankovich cycle as described above). For example, if the total methane volume in Titan's atmosphere and surface is not constant, then there might not have been enough methane available to fill the south seas at the periods of greatest solar insolation (i.e. heaviest rainfall). In one of the possible cases for Titan's methane abundance modeled by \citet{2012ApJ...749..159N}, the supply increases over time, a process that is probably driven by the brightening of the Sun over time \citep{Ribas2005} warming Titan's crust \citep{2012ApJ...749..159N}. In such a scenario, there may not have been enough liquid to fill the north pole lakes and seas, the equatorial basins, and the south polar basins. \\

\subsection{Equatorial Basins}

In light of the discussion above, what does the presence of 5-$\mu$m-bright material imply for the histories of Tui and Hotei? Based on Earth analogs, we assume that such a large areal extent of evaporite lying at the bottom of a basin is more consistent with a long-lived body of liquid, where a saturated solution has been in a closed system long enough to undergo a few cycles of evaporite formation to deposit the material we observe today. While Xanadu is a neighboring, low-lying region, its paucity of 5-$\mu$m-bright material is consistent with the requirement that liquid systems be sufficiently closed to form evaporite (at least on the order of the large scale deposits observable by VIMS): Xanadu is riddled with channels \citep{2009GeoRL..3622203B} that (probably) drain liquid into Tui and Hotei. \citet{Burr2013morph} discuss a variety of possible flow mechanisms that may be happening in Xanadu based on the observed fluvial features.\\

Assuming evaporite did form at the south pole and has subsequently been buried, why haven't the deposits covering Tui and Hotei suffered the same fate? As noted by \citet{2012Icar..221..768S}, the haze production rates of \citet{2003P&SS...51..963R,2004Icar..170..443R,2006Sci...311..201R,2009P&SS...57.1857W} span a range of values as do possible particle densities. If we use the lower estimate of the deposition rate of atmospheric particles to be about 0.1 $\mu$m per Titan year, it would take a few thousand Earth years to build up a layer of atmospheric particles on the surface of Tui and Hotei thick enough to mask the signal from any 5-$\mu$m-bright material. However, with higher estimates of haze production, on the order of 1 $\mu$m per Titan year, deposition would sufficiently build up after only a few hundred Earth years. Either Tui and Hotei were filled more recently than this (and thus also more recently than the proposed south polar seas) or the basins experience enough rainfall to dissolve and reprecipitate the evaporite deposits on timescales comparable to those predicted for equatorial storms. However, there is as yet no observational evidence for rainfall reaching Tui and Hotei. \\

It is also unlikely that a single storm of the present climate could have created the amount of liquid necessary to create Tui and Hotei's evaporites. The evaporites extend to the RADAR-identified boundaries of Tui and Hotei, thus the basins would have had to be completely filled to deposit evaporite at those farthest edges. According to the calculations of Section \ref{tuihote}, we show that this is a sizable volume of liquid. Comparisons between possible flooding mechanisms on Earth, Mars, and Titan led \cite{2010GPC....70....5B} to conclude that monsoonal flooding of the equatorial region (and thus the resulting carving of fluvial features) could be linked to a methane cycle. Periods of heavy rainfall would deplete the atmosphere, creating a more arid climate that would slowly be replenished until saturation in the atmosphere triggered another monsoon event. Pluvial events like these are responsible for some of the largest salt pans on Earth: Devil's Golf Course (Arizona, USA), Bonneville Salt Flats (Utah, USA), and the Lake Eyre (South Australia, Australia).\\

\subsection{Seasonality}

The most likely candidates for evaporitic compounds on Titan are the hydrocarbons that originate in the atmosphere that have been dissolved in the surface liquid. Titan's surface liquid is probably a complex mixture of methane, ethane, nitrogen, and small amounts of alkanes (e.g. propane) \citep{2013Icar..226.1431C,2013GeCoA.115..217G} (but see \citet{LigeiaDepth}). Experimental studies \citep{2012GeoRL..3923203L} and thermodynamic models \citep{2013GeCoA.115..217G} indicate that organic molecules have a higher solubility with liquid ethane than methane. While this affects the composition of the solution itself, it has less influence on the actual formation of evaporite. Evaporite can only form if the solution is saturated. Thus, regardless of the solubilities of the individual constituent solvents, solute will deposit when one evaporates. \\

The composition of the solution filling the lakes does raise questions concerning the seasonality of evaporation and evaporite formation. Methane is thought to be the only compound volatile enough to undergo evaporation over reasonable timescales; ethane has been proposed to be volatile on larger-than-seasonal timescales. Hence, if evaporation were the sole mechanism for evacuating both methane and ethane (as we expect for evaporite-bottomed lakes), the lake desiccation processes would take longer than methane evaporation alone. Alternatively, if an over-saturated, ethane-dominated lake were emptied of liquid via a regolith porous enough for the solution but not the undissolved solute to percolate through, then evaporite formed when the methane evaporated could be left behind. It is also possible to envision scenarios where different liquid-removing mechanisms have been dominant at different times.  Infiltration could drain out some of the liquid and evaporation became the dominant mechanism after the subsurface liquid reaches the water-table, for example.\\

Thus, the evaporite formation timescale could be complex: formation associated with the evaporation of methane occurring seasonally and evaporation of the entire liquid body's contents occurring over large enough timescales for ethane to evaporate, with allowances for rain influx. Presumably, such mechanisms would work faster with shallower bodies of liquid, such as those identified by \citet{evaporite}, where the local timescales of evaporation would contribute to the observed dichotomy in surface liquid distribution on Titan. In the case that evaporation is the only liquid removal process in the lifetime of the lake, seasonal exchange of methane alone, as discussed above, is \emph{not} enough to explain the observation of dry lakebeds with 5-$\mu$m-bright bottoms. \\

\section{Conclusion}  
Operating under the hypothesis put forth by \citet{evaporite} that coherent, 5-$\mu$m-bright signatures on Titan's surface are evaporitic in origin, we map the global distribution of 5-$\mu$m-bright material and thus present newly-identified evaporite candidates. These evaporite candidates cover a little over 1\% of Titan's surface area and are mostly concentrated in Tui and Hotei Regiones (proposed fossil seas of the equatorial region) and at the north pole (around the giant seas and dry, filled, or wetted lakebeds). Unlike those at their northern counterparts, the south polar basins are noticeably void of 5-$\mu$m-bright material, prompting the following possible explanations: evaporite deposits at the south pole have been subsequently covered or removed, or conditions were never suitable for evaporite formation. In these two scenarios, Tui and Hotei were then either filled more recently than the south polar basins or the processes responsible for making evaporite in the north pole also took place at the equatorial basins (but not at the south). As climate models seek to address the currently observed asymmetry in surface liquid distribution between the north and south polar regions, we encourage the consideration of each of these scenarios and their implications, including the more active role of the equatorial region implied by the evaporitic evidence for Tui and Hotei having once been filled seas. \\

Without the benefit of a lander, Titan's surface composition must be constrained through other, less direct means. We present this distribution of evaporite candidates as the first in a series of works investigating the 5-$\mu$m-bright spectral unit. Having now cataloged the total and best occurrences of each isolated deposit, we will next be able to construct spectra of high enough signal-to-noise to facilitate inter-flyby comparison. In particular, we will look for the absorption feature at 4.92 $\mu$m exhibited by Tui and Hotei in the spectra of other evaporite candidates \citep{McCord2008}. Especially in comparison to the spectra of 5-$\mu$m-bright deposits observed on the shores of filled lakes or dried lake beds, a sharing of this absorption feature would strengthen the case of the 5-$\mu$m-bright unit as diagnostic of material of evaporitic-origin. 

\section{Acknowledgements}
The authors would like to thank an anonymous reviewer for constructive suggestions to the manuscript. This work was supported by NASA Cassini Data Analysts and Participating Scientists (CDAPS) grant \#NNX12AC28G to JWB. CS acknowledges support from the NASA Astrobiology Institute. Part of this work was conducted at JPL/Caltech under contract with NASA.



  \bibliographystyle{elsarticle-harv} 
\bibliography{evaporite.global.distribution.bbl}

\begin{thebibliography}{94}
\expandafter\ifx\csname natexlab\endcsname\relax\def\natexlab#1{#1}\fi
\expandafter\ifx\csname url\endcsname\relax
  \def\url#1{\texttt{#1}}\fi
\expandafter\ifx\csname urlprefix\endcsname\relax\def\urlprefix{URL }\fi

\bibitem[{{Aharonson} et~al.(2009){Aharonson}, {Hayes}, {Lunine}, {Lorenz},
  {Allison}, and {Elachi}}]{2009NatGe...2..851A}
{Aharonson}, O., {Hayes}, A.~G., {Lunine}, J.~I., {Lorenz}, R.~D., {Allison},
  M.~D., {Elachi}, C., Dec. 2009. {An asymmetric distribution of lakes on Titan
  as a possible consequence of orbital forcing}. Nature Geoscience 2, 851--854.

\bibitem[{{Barnes} et~al.(2011){Barnes}, {Bow}, {Schwartz}, {Brown},
  {Soderblom}, {Hayes}, {Vixie}, {Le Mou{\'e}lic}, {Rodriguez}, {Sotin},
  {Jaumann}, {Stephan}, {Soderblom}, {Clark}, {Buratti}, {Baines}, and
  {Nicholson}}]{evaporite}
{Barnes}, J.~W., {Bow}, J., {Schwartz}, J., {Brown}, R.~H., {Soderblom}, J.~M.,
  {Hayes}, A.~G., {Vixie}, G., {Le Mou{\'e}lic}, S., {Rodriguez}, S., {Sotin},
  C., {Jaumann}, R., {Stephan}, K., {Soderblom}, L.~A., {Clark}, R.~N.,
  {Buratti}, B.~J., {Baines}, K.~H., {Nicholson}, P.~D., Nov. 2011. {Organic
  sedimentary deposits in Titan's dry lakebeds: Probable evaporite}. Icarus
  216, 136--140.

\bibitem[{{Barnes} et~al.(2006){Barnes}, {Brown}, {Radebaugh}, {Buratti},
  {Sotin}, {Le Mouelic}, {Rodriguez}, {Turtle}, {Perry}, {Clark}, {Baines}, and
  {Nicholson}}]{westTui}
{Barnes}, J.~W., {Brown}, R.~H., {Radebaugh}, J., {Buratti}, B.~J., {Sotin},
  C., {Le Mouelic}, S., {Rodriguez}, S., {Turtle}, E.~P., {Perry}, J., {Clark},
  R., {Baines}, K.~H., {Nicholson}, P.~D., Aug. 2006. {Cassini observations of
  flow-like features in western Tui Regio, Titan}. GRL 33, L16204.

\bibitem[{{Barnes} et~al.(2009){Barnes}, {Brown}, {Soderblom}, {Soderblom},
  {Jaumann}, {Jackson}, {Le Mou{\'e}lic}, {Sotin}, {Buratti}, {Pitman},
  {Baines}, {Clark}, {Nicholson}, {Turtle}, and {Perry}}]{BarnesOntario}
{Barnes}, J.~W., {Brown}, R.~H., {Soderblom}, J.~M., {Soderblom}, L.~A.,
  {Jaumann}, R., {Jackson}, B., {Le Mou{\'e}lic}, S., {Sotin}, C., {Buratti},
  B.~J., {Pitman}, K.~M., {Baines}, K.~H., {Clark}, R.~N., {Nicholson}, P.~D.,
  {Turtle}, E.~P., {Perry}, J., May 2009. {Shoreline features of Titan's
  Ontario Lacus from Cassini/VIMS observations}. Icarus 201, 217--225.

\bibitem[{{Barnes} et~al.(2007{\natexlab{a}}){Barnes}, {Brown}, {Soderblom},
  {Buratti}, {Sotin}, {Rodriguez}, {Le Mou{\`e}lic}, {Baines}, {Clark}, and
  {Nicholson}}]{2007mapping}
{Barnes}, J.~W., {Brown}, R.~H., {Soderblom}, L., {Buratti}, B.~J., {Sotin},
  C., {Rodriguez}, S., {Le Mou{\`e}lic}, S., {Baines}, K.~H., {Clark}, R.,
  {Nicholson}, P., Jan. 2007{\natexlab{a}}. {Global-scale surface spectral
  variations on Titan seen from Cassini/VIMS}. Icarus 186, 242--258.

\bibitem[{{Barnes} et~al.(2008){Barnes}, {Brown}, {Soderblom}, {Sotin}, {Le
  Mou{\`e}lic}, {Rodriguez}, {Jaumann}, {Beyer}, {Buratti}, {Pitman}, {Baines},
  {Clark}, and {Nicholson}}]{T20dunes}
{Barnes}, J.~W., {Brown}, R.~H., {Soderblom}, L., {Sotin}, C., {Le
  Mou{\`e}lic}, S., {Rodriguez}, S., {Jaumann}, R., {Beyer}, R.~A., {Buratti},
  B.~J., {Pitman}, K., {Baines}, K.~H., {Clark}, R., {Nicholson}, P., May 2008.
  {Spectroscopy, morphometry, and photoclinometry of Titan's dunefields from
  Cassini/VIMS}. Icarus 195, 400--414, doi:10.1016/j.icarus.2007.12.006.

\bibitem[{{Barnes} et~al.(2005){Barnes}, {Brown}, {Turtle}, {McEwen}, {Lorenz},
  {Janssen}, {Schaller}, {Brown}, {Buratti}, {Sotin}, {Griffith}, {Clark},
  {Perry}, {Fussner}, {Barbara}, {West}, {Elachi}, {Bouchez}, {Roe}, {Baines},
  {Bellucci}, {Bibring}, {Capaccioni}, {Cerroni}, {Combes}, {Coradini},
  {Cruikshank}, {Drossart}, {Formisano}, {Jaumann}, {Langevin}, {Matson},
  {McCord}, {Nicholson}, and {Sicardy}}]{2005Sci...310...92B}
{Barnes}, J.~W., {Brown}, R.~H., {Turtle}, E.~P., {McEwen}, A.~S., {Lorenz},
  R.~D., {Janssen}, M., {Schaller}, E.~L., {Brown}, M.~E., {Buratti}, B.~J.,
  {Sotin}, C., {Griffith}, C., {Clark}, R., {Perry}, J., {Fussner}, S.,
  {Barbara}, J., {West}, R., {Elachi}, C., {Bouchez}, A.~H., {Roe}, H.~G.,
  {Baines}, K.~H., {Bellucci}, G., {Bibring}, J.-P., {Capaccioni}, F.,
  {Cerroni}, P., {Combes}, M., {Coradini}, A., {Cruikshank}, D.~P., {Drossart},
  P., {Formisano}, V., {Jaumann}, R., {Langevin}, Y., {Matson}, D.~L.,
  {McCord}, T.~B., {Nicholson}, P.~D., {Sicardy}, B., Oct. 2005. {A
  5-Micron-Bright Spot on Titan: Evidence for Surface Diversity}. Science 310,
  92--95.

\bibitem[{{Barnes} et~al.(2013){Barnes}, {Buratti}, {Turtle}, {Bow}, {Dalba},
  {Perry}, {Brown}, {Rodriguez}, {Le Mou\'elic}, {Baines}, {Sotin}, {Lorenz},
  {Malaska}, McCord, {Clark}, {Jaumann}, {Hayne}, {Nicholson}, {Soderblom}, and
  {Soderblom}}]{PrecipSurfaceBrightening}
{Barnes}, J.~W., {Buratti}, B.~J., {Turtle}, E.~P., {Bow}, J., {Dalba}, P.~A.,
  {Perry}, J.~E., {Brown}, R.~H., {Rodriguez}, S., {Le Mou\'elic}, S.,
  {Baines}, K.~H., {Sotin}, C., {Lorenz}, R.~D., {Malaska}, M.~J., McCord,
  T.~B., {Clark}, R.~N., {Jaumann}, R., {Hayne}, P.~O., {Nicholson}, P.~D.,
  {Soderblom}, J.~M., {Soderblom}, L.~A., Jan. 2013. {Precipitation-Induced
  Surface Brightenings Seen on Titan by Cassini VIMS and ISS}. Planetary
  Science 2.

\bibitem[{{Barnes} et~al.(2007{\natexlab{b}}){Barnes}, {Radebaugh}, {Brown},
  {Wall}, {Soderblom}, {Lunine}, {Burr}, {Sotin}, {Le Mou{\'e}lic},
  {Rodriguez}, {Buratti}, {Clark}, {Baines}, {Jaumann}, {Nicholson}, {Kirk},
  {Lopes}, {Lorenz}, {Mitchell}, and {Wood}}]{BarnesMountains}
{Barnes}, J.~W., {Radebaugh}, J., {Brown}, R.~H., {Wall}, S., {Soderblom}, L.,
  {Lunine}, J., {Burr}, D., {Sotin}, C., {Le Mou{\'e}lic}, S., {Rodriguez}, S.,
  {Buratti}, B.~J., {Clark}, R., {Baines}, K.~H., {Jaumann}, R., {Nicholson},
  P.~D., {Kirk}, R.~L., {Lopes}, R., {Lorenz}, R.~D., {Mitchell}, K., {Wood},
  C.~A., Nov. 2007{\natexlab{b}}. {Near-infrared spectral mapping of Titan's
  mountains and channels}. JGR Planets 112, 11006.

\bibitem[{{Barth}(2010)}]{BarthRidgeClouds}
{Barth}, E.~L., Nov. 2010. {Cloud formation along mountain ridges on Titan}.
  P\&SS 58, 1740--1747.

\bibitem[{{Brown} et~al.(2010){Brown}, {Roberts}, and
  {Schaller}}]{2010Icar..205..571B}
{Brown}, M.~E., {Roberts}, J.~E., {Schaller}, E.~L., Feb. 2010. {Clouds on
  Titan during the Cassini prime mission: A complete analysis of the VIMS
  data}. Icarus 205, 571--580.

\bibitem[{{Brown} et~al.(2009){Brown}, {Smith}, {Chen}, and
  {{\'A}d{\'a}mkovics}}]{2009ApJ...706L.110B}
{Brown}, M.~E., {Smith}, A.~L., {Chen}, C., {{\'A}d{\'a}mkovics}, M., Nov.
  2009. {Discovery of Fog at the South Pole of Titan}. ApJ-L 706, L110--L113.

\bibitem[{{Brown} et~al.(2008){Brown}, {Soderblom}, {Soderblom}, {Clark},
  {Jaumann}, {Barnes}, {Sotin}, {Buratti}, {Baines}, and
  {Nicholson}}]{T38.ethane}
{Brown}, R.~H., {Soderblom}, L.~A., {Soderblom}, J.~M., {Clark}, R.~N.,
  {Jaumann}, R., {Barnes}, J.~W., {Sotin}, C., {Buratti}, B., {Baines}, K.~H.,
  {Nicholson}, P.~D., Jul. 2008. {The identification of liquid ethane in
  Titan's Ontario Lacus}. Nature 454, 607--610.

\bibitem[{{Burr}(2010)}]{2010GPC....70....5B}
{Burr}, D.~M., Feb. 2010. {Palaeoflood-generating mechanisms on Earth, Mars,
  and Titan}. Global and Planetary Change 70, 5--13.

\bibitem[{Burr et~al.(2013)Burr, Drummond, Cartwright, Black, and
  Perron}]{Burr2013742}
Burr, D.~M., Drummond, S.~A., Cartwright, R., Black, B.~A., Perron, J.~T.,
  2013. Morphology of fluvial networks on titan: Evidence for structural
  control. Icarus 226~(1), 742 -- 759.
\newline\urlprefix\url{http://www.sciencedirect.com/science/article/pii/S0019103513002728}

\bibitem[{{Burr} et~al.(2009){Burr}, {Jacobsen}, {Roth}, {Phillips},
  {Mitchell}, and {Viola}}]{2009GeoRL..3622203B}
{Burr}, D.~M., {Jacobsen}, R.~E., {Roth}, D.~L., {Phillips}, C.~B., {Mitchell},
  K.~L., {Viola}, D., Nov. 2009. {Fluvial network analysis on Titan: Evidence
  for subsurface structures and west-to-east wind flow, southwestern Xanadu}.
  GRL 36, 22203.

\bibitem[{{Burr} et~al.(2013){Burr}, {Perron}, {Lamb}, {Irwin}, {Collins},
  {Howard}, {Sklar}, {Moore}, , {Adamkovics}, {Baker}, and
  {Black}}]{Burr2013morph}
{Burr}, D.~M., {Perron}, T., {Lamb}, M.~P., {Irwin}, R.~P., {Collins}, G.~C.,
  {Howard}, A.~D., {Sklar}, L.~S., {Moore}, J.~M., , {Adamkovics}, M., {Baker},
  V.~R.~{Drummond}, S., {Black}, B.~A., 2013. {Fluvial features on Titan:
  Insights from morphology and modeling}. Bulletin of the Geological Society of
  America 125, 229 -- 321.
\newline\urlprefix\url{http://www.sciencedirect.com/science/article/pii/S0019103513002728}

\bibitem[{{Cook} et~al.(submitted){Cook}, {Barnes}, {Kattenhorn}, {Radebaugh},
  and {Beuthe}}]{CookMtns}
{Cook}, C., {Barnes}, J.~W., {Kattenhorn}, S.~A., {Radebaugh}, J., {Beuthe},
  M., submitted. {Evidence for Global Contraction on Titan from Patterns of
  Tectonism}. JGRP.

\bibitem[{{Cordier} et~al.(2013){Cordier}, {Barnes}, and
  {Ferreira}}]{2013Icar..226.1431C}
{Cordier}, D., {Barnes}, J.~W., {Ferreira}, A.~G., Nov. 2013. {On the chemical
  composition of Titan's dry lakebed evaporites}. Icarus 226, 1431--1437.

\bibitem[{{Cordier} et~al.(2009){Cordier}, {Mousis}, {Lunine}, {Lavvas}, and
  {Vuitton}}]{2009ApJ...707L.128C}
{Cordier}, D., {Mousis}, O., {Lunine}, J.~I., {Lavvas}, P., {Vuitton}, V., Dec.
  2009. {An Estimate of the Chemical Composition of Titan's Lakes}. ApJ-L 707,
  L128--L131.

\bibitem[{{Cornet} et~al.(2012{\natexlab{a}}){Cornet}, {Bourgeois}, {Le
  Mou{\'e}lic}, {Rodriguez}, {Lopez Gonzalez}, {Sotin}, {Tobie}, {Fleurant},
  {Barnes}, {Brown}, {Baines}, {Buratti}, {Clark}, and
  {Nicholson}}]{Ontario.Etosha}
{Cornet}, T., {Bourgeois}, O., {Le Mou{\'e}lic}, S., {Rodriguez}, S., {Lopez
  Gonzalez}, T., {Sotin}, C., {Tobie}, G., {Fleurant}, C., {Barnes}, J.~W.,
  {Brown}, R.~H., {Baines}, K.~H., {Buratti}, B.~J., {Clark}, R.~N.,
  {Nicholson}, P.~D., Apr. 2012{\natexlab{a}}. {Geomorphological significance
  of Ontario Lacus on Titan: Integrated interpretation of Cassini VIMS, ISS and
  RADAR data and comparison with the Etosha Pan (Namibia)}. Icarus 218,
  788--806.

\bibitem[{{Cornet} et~al.(2012{\natexlab{b}}){Cornet}, {Bourgeois}, {Le
  Mou{\'e}lic}, {Rodriguez}, {Sotin}, {Barnes}, {Brown}, {Baines}, {Buratti},
  {Clark}, and {Nicholson}}]{Ontario.nochange}
{Cornet}, T., {Bourgeois}, O., {Le Mou{\'e}lic}, S., {Rodriguez}, S., {Sotin},
  C., {Barnes}, J.~W., {Brown}, R.~H., {Baines}, K.~H., {Buratti}, B.~J.,
  {Clark}, R.~N., {Nicholson}, P.~D., Jul. 2012{\natexlab{b}}. {Edge detection
  applied to Cassini images reveals no measurable displacement of Ontario
  Lacus' margin between 2005 and 2010}. JGR Planets 117, 7005.

\bibitem[{{Elachi} et~al.(2006){Elachi}, {Wall}, {Janssen}, {Stofan}, {Lopes},
  {Kirk}, {Lorenz}, {Lunine}, {Paganelli}, {Soderblom}, {Wood}, {Wye},
  {Zebker}, {Anderson}, {Ostro}, {Allison}, {Boehmer}, {Callahan}, {Encrenaz},
  {Flamini}, {Francescetti}, {Gim}, {Hamilton}, {Hensley}, {Johnson},
  {Kelleher}, {Muhleman}, {Picardi}, {Posa}, {Roth}, {Seu}, {Shaffer},
  {Stiles}, {Vetrella}, and {West}}]{2006Natur.441..709E}
{Elachi}, C., {Wall}, S., {Janssen}, M., {Stofan}, E., {Lopes}, R., {Kirk}, R.,
  {Lorenz}, R., {Lunine}, J., {Paganelli}, F., {Soderblom}, L., {Wood}, C.,
  {Wye}, L., {Zebker}, H., {Anderson}, Y., {Ostro}, S., {Allison}, M.,
  {Boehmer}, R., {Callahan}, P., {Encrenaz}, P., {Flamini}, E., {Francescetti},
  G., {Gim}, Y., {Hamilton}, G., {Hensley}, S., {Johnson}, W., {Kelleher}, K.,
  {Muhleman}, D., {Picardi}, G., {Posa}, F., {Roth}, L., {Seu}, R., {Shaffer},
  S., {Stiles}, B., {Vetrella}, S., {West}, R., Jun. 2006. {Titan Radar Mapper
  observations from Cassini's T3 fly-by}. Nature 441, 709--713.

\bibitem[{{Flasar}(1983)}]{1983Sci...221...55F}
{Flasar}, F.~M., Jul. 1983. {Oceans on Titan?} Science 221, 55--57.

\bibitem[{{Glein} and {Shock}(2013)}]{2013GeCoA.115..217G}
{Glein}, C.~R., {Shock}, E.~L., Aug. 2013. {A geochemical model of non-ideal
  solutions in the methane-ethane-propane-nitrogen-acetylene system on Titan}.
  Geochimica et Cosmochimica Acta 115, 217--240.

\bibitem[{Griffith(2009)}]{Griffith09}
Griffith, C.~A., 2009. {Storms, polar deposits and the methane cycle in Titan's
  atmosphere}. Philosophical Transactions of the Royal Society A: Mathematical,
  Physical and Engineering Sciences 367~(1889), 713--728.
\newline\urlprefix\url{http://rsta.royalsocietypublishing.org/content/367/1889/713.abstract}

\bibitem[{{Griffith} et~al.(2012){Griffith}, {Lora}, {Turner}, {Penteado},
  {Brown}, {Tomasko}, {Doose}, and {See}}]{2012Natur.486..237G}
{Griffith}, C.~A., {Lora}, J.~M., {Turner}, J., {Penteado}, P.~F., {Brown},
  R.~H., {Tomasko}, M.~G., {Doose}, L., {See}, C., Jun. 2012. {Possible
  tropical lakes on Titan from observations of dark terrain}. Nature 486,
  237--239.

\bibitem[{{Griffith} et~al.(1998){Griffith}, {Owen}, {Miller}, and
  {Geballe}}]{1998Natur.395..575G}
{Griffith}, C.~A., {Owen}, T., {Miller}, G.~A., {Geballe}, T., 1998. {Transient
  clouds in Titan's lower atmosphere.} Nature 395, 575--578.

\bibitem[{{Griffith} et~al.(2005){Griffith}, {Penteado}, {Baines}, {Drossart},
  {Barnes}, {Bellucci}, {Bibring}, {Brown}, {Buratti}, {Capaccioni}, {Cerroni},
  {Clark}, {Combes}, {Coradini}, {Cruikshank}, {Formisano}, {Jaumann},
  {Langevin}, {Matson}, {McCord}, {Mennella}, {Nelson}, {Nicholson}, {Sicardy},
  {Sotin}, {Soderblom}, and {Kursinski}}]{2005Sci...310..474G}
{Griffith}, C.~A., {Penteado}, P., {Baines}, K., {Drossart}, P., {Barnes}, J.,
  {Bellucci}, G., {Bibring}, J., {Brown}, R., {Buratti}, B., {Capaccioni}, F.,
  {Cerroni}, P., {Clark}, R., {Combes}, M., {Coradini}, A., {Cruikshank}, D.,
  {Formisano}, V., {Jaumann}, R., {Langevin}, Y., {Matson}, D., {McCord}, T.,
  {Mennella}, V., {Nelson}, R., {Nicholson}, P., {Sicardy}, B., {Sotin}, C.,
  {Soderblom}, L.~A., {Kursinski}, R., Oct. 2005. {The Evolution of Titan's
  Mid-Latitude Clouds}. Science 310, 474--477.

\bibitem[{{Griffith} et~al.(2006){Griffith}, {Penteado}, {Rannou}, {Brown},
  {Boudon}, {Baines}, {Clark}, {Drossart}, {Buratti}, {Nicholson}, {McKay},
  {Coustenis}, {Negrao}, and {Jaumann}}]{2006Sci...313.1620G}
{Griffith}, C.~A., {Penteado}, P., {Rannou}, P., {Brown}, R., {Boudon}, V.,
  {Baines}, K.~H., {Clark}, R., {Drossart}, P., {Buratti}, B., {Nicholson}, P.,
  {McKay}, C.~P., {Coustenis}, A., {Negrao}, A., {Jaumann}, R., Sep. 2006.
  {Evidence for a Polar Ethane Cloud on Titan}. Science 313, 1620--1622.

\bibitem[{{Hayes} et~al.(2008){Hayes}, {Aharonson}, {Callahan}, {Elachi},
  {Gim}, {Kirk}, {Lewis}, {Lopes}, {Lorenz}, {Lunine}, {Mitchell}, {Mitri},
  {Stofan}, and {Wall}}]{Hayes2008}
{Hayes}, A., {Aharonson}, O., {Callahan}, P., {Elachi}, C., {Gim}, Y., {Kirk},
  R., {Lewis}, K., {Lopes}, R., {Lorenz}, R., {Lunine}, J., {Mitchell}, K.,
  {Mitri}, G., {Stofan}, E., {Wall}, S., May 2008. {Hydrocarbon lakes on Titan:
  Distribution and interaction with a porous regolith}. Geophysical Research
  Letters 35, L9204.

\bibitem[{{Hayes} et~al.(2011){Hayes}, {Aharonson}, {Lunine}, {Kirk}, {Zebker},
  {Wye}, {Lorenz}, {Turtle}, {Paillou}, {Mitri}, {Wall}, {Stofan}, {Mitchell},
  {Elachi}, and {the Cassini RADAR Team}}]{Hayes2011}
{Hayes}, A.~G., {Aharonson}, O., {Lunine}, J.~I., {Kirk}, R.~L., {Zebker},
  H.~A., {Wye}, L.~C., {Lorenz}, R.~D., {Turtle}, E.~P., {Paillou}, P.,
  {Mitri}, G., {Wall}, S.~D., {Stofan}, E.~R., {Mitchell}, K.~L., {Elachi}, C.,
  {the Cassini RADAR Team}, Jan. 2011. {Transient surface liquid in Titan's
  polar regions from Cassini}. Icarus 211, 655--671.

\bibitem[{{Hayes} et~al.(2014){Hayes}, {Michaelides}, {Turtle}, {Barnes},
  {Soderblom}, {Masrtogiuseppe}, {Lorenz}, {Kirk}, and
  {Lunine}}]{2014LPI....45.2341H}
{Hayes}, A.~G., {Michaelides}, R.~J., {Turtle}, E.~P., {Barnes}, J.~W.,
  {Soderblom}, J.~M., {Masrtogiuseppe}, M., {Lorenz}, R.~D., {Kirk}, R.~L.,
  {Lunine}, J.~I., Mar. 2014. {The Distribution and Volume of Titan's
  Hydrocarbon Lakes and Seas}. In: Lunar and Planetary Science Conference.
  Vol.~45 of Lunar and Planetary Science Conference. p. 2341.

\bibitem[{{Hayes} et~al.(2010){Hayes}, {Wolf}, {Aharonson}, {Zebker}, {Lorenz},
  {Kirk}, {Paillou}, {Lunine}, {Wye}, {Callahan}, {Wall}, and
  {Elachi}}]{Hayes2010}
{Hayes}, A.~G., {Wolf}, A.~S., {Aharonson}, O., {Zebker}, H., {Lorenz}, R.,
  {Kirk}, R.~L., {Paillou}, P., {Lunine}, J., {Wye}, L., {Callahan}, P.,
  {Wall}, S., {Elachi}, C., Sep. 2010. {Bathymetry and absorptivity of Titan's
  Ontario Lacus}. Journal of Geophysical Research (Planets) 115, 9009.

\bibitem[{{Hirtzig} et~al.(2009){Hirtzig}, {Tokano}, {Rodriguez}, {Le
  Mou{\'e}lic}, and {Sotin}}]{2009A&ARv..17..105H}
{Hirtzig}, M., {Tokano}, T., {Rodriguez}, S., {Le Mou{\'e}lic}, S., {Sotin},
  C., Jun. 2009. {A review of Titan's atmospheric phenomena}. JGR Planets 17,
  105--147.

\bibitem[{{Iess} et~al.(2012){Iess}, {Jacobson}, {Ducci}, {Stevenson},
  {Lunine}, {Armstrong}, {Asmar}, {Racioppa}, {Rappaport}, and
  {Tortora}}]{2012Sci...337..457I}
{Iess}, L., {Jacobson}, R.~A., {Ducci}, M., {Stevenson}, D.~J., {Lunine},
  J.~I., {Armstrong}, J.~W., {Asmar}, S.~W., {Racioppa}, P., {Rappaport},
  N.~J., {Tortora}, P., Jul. 2012. {The Tides of Titan}. Science 337, 457--.

\bibitem[{{Jaumann} et~al.(2008){Jaumann}, {Brown}, {Stephan}, {Barnes},
  {Soderblom}, {Sotin}, {Le Mou{\'e}lic}, {Clark}, {Soderblom}, {Buratti},
  {Wagner}, {McCord}, {Rodriguez}, {Baines}, {Cruikshank}, {Nicholson},
  {Griffith}, {Langhans}, and {Lorenz}}]{Jaumann08}
{Jaumann}, R., {Brown}, R.~H., {Stephan}, K., {Barnes}, J.~W., {Soderblom},
  L.~A., {Sotin}, C., {Le Mou{\'e}lic}, S., {Clark}, R.~N., {Soderblom}, J.,
  {Buratti}, B.~J., {Wagner}, R., {McCord}, T.~B., {Rodriguez}, S., {Baines},
  K.~H., {Cruikshank}, D.~P., {Nicholson}, P.~D., {Griffith}, C.~A.,
  {Langhans}, M., {Lorenz}, R.~D., Oct. 2008. {Fluvial erosion and
  post-erosional processes on Titan}. Icarus 197, 526--538.

\bibitem[{{Langhans} et~al.(2012){Langhans}, {Jaumann}, {Stephan}, {Brown},
  {Buratti}, {Clark}, {Baines}, {Nicholson}, {Lorenz}, {Soderblom},
  {Soderblom}, {Sotin}, {Barnes}, and {Nelson}}]{2012P&SS...60...34L}
{Langhans}, M.~H., {Jaumann}, R., {Stephan}, K., {Brown}, R.~H., {Buratti},
  B.~J., {Clark}, R.~N., {Baines}, K.~H., {Nicholson}, P.~D., {Lorenz}, R.~D.,
  {Soderblom}, L.~A., {Soderblom}, J.~M., {Sotin}, C., {Barnes}, J.~W.,
  {Nelson}, R., Jan. 2012. {Titan's fluvial valleys: Morphology, distribution,
  and spectral properties}. P\&SS 60, 34--51.

\bibitem[{{Le Mou{\'e}lic} et~al.(2008){Le Mou{\'e}lic}, {Paillou}, {Janssen},
  {Barnes}, {Rodriguez}, {Sotin}, {Brown}, {Baines}, {Buratti}, {Clark},
  {Crapeau}, {Encrenaz}, {Jaumann}, {Geudtner}, {Paganelli}, {Soderblom},
  {Tobie}, and {Wall}}]{StephaneSinlap}
{Le Mou{\'e}lic}, S., {Paillou}, P., {Janssen}, M.~A., {Barnes}, J.~W.,
  {Rodriguez}, S., {Sotin}, C., {Brown}, R.~H., {Baines}, K.~H., {Buratti},
  B.~J., {Clark}, R.~N., {Crapeau}, M., {Encrenaz}, P.~J., {Jaumann}, R.,
  {Geudtner}, D., {Paganelli}, F., {Soderblom}, L., {Tobie}, G., {Wall}, S.,
  Apr. 2008. {Mapping and interpretation of Sinlap crater on Titan using
  Cassini VIMS and RADAR data}. JGR Planets 113, 4003.

\bibitem[{{Le Mou{\'e}lic} et~al.(2012){Le Mou{\'e}lic}, {Rannou}, {Rodriguez},
  {Sotin}, {Griffith}, {Le Corre}, {Barnes}, {Brown}, {Baines}, {Buratti},
  {Clark}, {Nicholson}, and {Tobie}}]{2012P&SS...60...86L}
{Le Mou{\'e}lic}, S., {Rannou}, P., {Rodriguez}, S., {Sotin}, C., {Griffith},
  C.~A., {Le Corre}, L., {Barnes}, J.~W., {Brown}, R.~H., {Baines}, K.~H.,
  {Buratti}, B.~J., {Clark}, R.~N., {Nicholson}, P.~D., {Tobie}, G., Jan. 2012.
  {Dissipation of Titan's north polar cloud at northern spring equinox}. P\&SS
  60, 86--92.

\bibitem[{{Lopes} et~al.(2013){Lopes}, {Kirk}, {Mitchell}, {Legall}, {Barnes},
  {Hayes}, {Kargel}, {Wye}, {Radebaugh}, {Stofan}, {Janssen}, {Neish}, {Wall},
  {Wood}, {Lunine}, and {Malaska}}]{2013JGRE..118..416L}
{Lopes}, R.~M.~C., {Kirk}, R.~L., {Mitchell}, K.~L., {Legall}, A., {Barnes},
  J.~W., {Hayes}, A., {Kargel}, J., {Wye}, L., {Radebaugh}, J., {Stofan},
  E.~R., {Janssen}, M.~A., {Neish}, C.~D., {Wall}, S.~D., {Wood}, C.~A.,
  {Lunine}, J.~I., {Malaska}, M.~J., Mar. 2013. {Cryovolcanism on Titan: New
  results from Cassini RADAR and VIMS}. JGR Planets 118, 416--435.

\bibitem[{{Lopes} et~al.(2010){Lopes}, {Stofan}, {Peckyno}, {Radebaugh},
  {Mitchell}, {Mitri}, {Wood}, {Kirk}, {Wall}, {Lunine}, {Hayes}, {Lorenz},
  {Farr}, {Wye}, {Craig}, {Ollerenshaw}, {Janssen}, {Legall}, {Paganelli},
  {West}, {Stiles}, {Callahan}, {Anderson}, {Valora}, {Soderblom}, and {Cassini
  RADAR Team}}]{Lopes2010}
{Lopes}, R.~M.~C., {Stofan}, E.~R., {Peckyno}, R., {Radebaugh}, J., {Mitchell},
  K.~L., {Mitri}, G., {Wood}, C.~A., {Kirk}, R.~L., {Wall}, S.~D., {Lunine},
  J.~I., {Hayes}, A., {Lorenz}, R., {Farr}, T., {Wye}, L., {Craig}, J.,
  {Ollerenshaw}, R.~J., {Janssen}, M., {Legall}, A., {Paganelli}, F., {West},
  R., {Stiles}, B., {Callahan}, P., {Anderson}, Y., {Valora}, P., {Soderblom},
  L., {Cassini RADAR Team}, Feb. 2010. {Distribution and interplay of geologic
  processes on Titan from Cassini radar data}. Icarus 205, 540--558.

\bibitem[{Lorenz et~al.(2014)Lorenz, Kirk, Hayes, Anderson, Lunine, Tokano,
  Turtle, Malaska, Soderblom, Lucas, Özgür Karatekin, and Wall}]{LorenzThroat}
Lorenz, R.~D., Kirk, R.~L., Hayes, A.~G., Anderson, Y.~Z., Lunine, J.~I.,
  Tokano, T., Turtle, E.~P., Malaska, M.~J., Soderblom, J.~M., Lucas, A., Özgür
  Karatekin, Wall, S.~D., 2014. A radar map of titan seas: Tidal dissipation
  and ocean mixing through the throat of kraken. Icarus 237~(0), 9 -- 15.
\newline\urlprefix\url{http://www.sciencedirect.com/science/article/pii/S0019103514001973}

\bibitem[{{Lorenz} et~al.(2008{\natexlab{a}}){Lorenz}, {Lopes}, {Paganelli},
  {Lunine}, {Kirk}, {Mitchell}, {Soderblom}, {Stofan}, {Ori}, {Myers},
  {Miyamoto}, {Radebaugh}, {Stiles}, {Wall}, and {Wood}}]{2008P&SS...56.1132L}
{Lorenz}, R.~D., {Lopes}, R.~M., {Paganelli}, F., {Lunine}, J.~I., {Kirk},
  R.~L., {Mitchell}, K.~L., {Soderblom}, L.~A., {Stofan}, E.~R., {Ori}, G.,
  {Myers}, M., {Miyamoto}, H., {Radebaugh}, J., {Stiles}, B., {Wall}, S.~D.,
  {Wood}, C.~A., Jun. 2008{\natexlab{a}}. {Fluvial channels on Titan: Initial
  Cassini RADAR observations}. P\&SS 56, 1132--1144.

\bibitem[{{Lorenz} and {Lunine}(1996)}]{1996Icar..122...79L}
{Lorenz}, R.~D., {Lunine}, J.~I., Jul. 1996. {Erosion on Titan: Past and
  Present}. Icarus 122, 79--91.

\bibitem[{{Lorenz} and {Lunine}(2005)}]{2005P&SS...53..557L}
{Lorenz}, R.~D., {Lunine}, J.~I., Apr. 2005. {Titan's surface before Cassini}.
  P\&SS 53, 557--576.

\bibitem[{{Lorenz} et~al.(2008{\natexlab{b}}){Lorenz}, {Mitchell}, {Kirk},
  {Hayes}, {Aharonson}, {Zebker}, {Paillou}, {Radebaugh}, {Lunine}, {Janssen},
  {Wall}, {Lopes}, {Stiles}, {Ostro}, {Mitri}, and
  {Stofan}}]{LorenzSurfaceInventory}
{Lorenz}, R.~D., {Mitchell}, K.~L., {Kirk}, R.~L., {Hayes}, A.~G., {Aharonson},
  O., {Zebker}, H.~A., {Paillou}, P., {Radebaugh}, J., {Lunine}, J.~I.,
  {Janssen}, M.~A., {Wall}, S.~D., {Lopes}, R.~M., {Stiles}, B., {Ostro}, S.,
  {Mitri}, G., {Stofan}, E.~R., Jan. 2008{\natexlab{b}}. {Titan's inventory of
  organic surface materials}. GRL 35, 2206.

\bibitem[{{Lorenz} and {Radebaugh}(2009)}]{2009GeoRL..36.3202L}
{Lorenz}, R.~D., {Radebaugh}, J., Feb. 2009. {Global pattern of Titan's dunes:
  Radar survey from the Cassini prime mission}. GRL 36, 3202.

\bibitem[{{Lorenz} et~al.(2013){Lorenz}, {Stiles}, {Aharonson}, {Lucas},
  {Hayes}, {Kirk}, {Zebker}, {Turtle}, {Neish}, {Stofan}, and
  {Barnes}}]{LorenzTopo}
{Lorenz}, R.~D., {Stiles}, B.~W., {Aharonson}, O., {Lucas}, A., {Hayes}, A.~G.,
  {Kirk}, R.~L., {Zebker}, H.~A., {Turtle}, E.~P., {Neish}, C.~D., {Stofan},
  E.~R., {Barnes}, J.~W., Jul. 2013. {A global topographic map of Titan}.
  Icarus 225, 367--377.

\bibitem[{{Lunine} et~al.(2008){Lunine}, {Elachi}, {Wall}, {Janssen},
  {Allison}, {Anderson}, {Boehmer}, {Callahan}, {Encrenaz}, {Flamini},
  {Franceschetti}, {Gim}, {Hamilton}, {Hensley}, {Johnson}, {Kelleher}, {Kirk},
  {Lopes}, {Lorenz}, {Muhleman}, {Orosei}, {Ostro}, {Paganelli}, {Paillou},
  {Picardi}, {Posa}, {Radebaugh}, {Roth}, {Seu}, {Shaffer}, {Soderblom},
  {Stiles}, {Stofan}, {Vetrella}, {West}, {Wood}, {Wye}, {Zebker}, {Alberti},
  {Karkoschka}, {Rizk}, {McFarlane}, {See}, and
  {Kazeminejad}}]{2008Icar..195..415L}
{Lunine}, J.~I., {Elachi}, C., {Wall}, S.~D., {Janssen}, M.~A., {Allison},
  M.~D., {Anderson}, Y., {Boehmer}, R., {Callahan}, P., {Encrenaz}, P.,
  {Flamini}, E., {Franceschetti}, G., {Gim}, Y., {Hamilton}, G., {Hensley}, S.,
  {Johnson}, W.~T.~K., {Kelleher}, K., {Kirk}, R.~L., {Lopes}, R.~M., {Lorenz},
  R., {Muhleman}, D.~O., {Orosei}, R., {Ostro}, S.~J., {Paganelli}, F.,
  {Paillou}, P., {Picardi}, G., {Posa}, F., {Radebaugh}, J., {Roth}, L.~E.,
  {Seu}, R., {Shaffer}, S., {Soderblom}, L.~A., {Stiles}, B., {Stofan}, E.~R.,
  {Vetrella}, S., {West}, R., {Wood}, C.~A., {Wye}, L., {Zebker}, H.,
  {Alberti}, G., {Karkoschka}, E., {Rizk}, B., {McFarlane}, E., {See}, C.,
  {Kazeminejad}, B., May 2008. {Titan's diverse landscapes as evidenced by
  Cassini RADAR's third and fourth looks at Titan}. Icarus 195, 415--433.

\bibitem[{{Lunine} et~al.(1983){Lunine}, {Stevenson}, and
  {Yung}}]{1983Sci...222.1229L}
{Lunine}, J.~I., {Stevenson}, D.~J., {Yung}, Y.~L., Dec. 1983. {Ethane ocean on
  Titan}. Science 222, 1229--1230.

\bibitem[{{Luspay-Kuti} et~al.(2012){Luspay-Kuti}, {Chevrier}, {Wasiak}, {Roe},
  {Welivitiya}, {Cornet}, {Singh}, and {Rivera-Valentin}}]{2012GeoRL..3923203L}
{Luspay-Kuti}, A., {Chevrier}, V.~F., {Wasiak}, F.~C., {Roe}, L.~A.,
  {Welivitiya}, W.~D.~D.~P., {Cornet}, T., {Singh}, S., {Rivera-Valentin},
  E.~G., Dec. 2012. {Experimental simulations of CH$_{4}$ evaporation on
  Titan}. GRL 39, 23203.

\bibitem[{Malaska and Hodyss(2014)}]{Malaska2014}
Malaska, M.~J., Hodyss, R., 2014. Dissolution of benzene, naphthalene, and
  biphenyl in a simulated titan lake. Icarus~(0), --.
\newline\urlprefix\url{http://www.sciencedirect.com/science/article/pii/S0019103514004072}

\bibitem[{{Mandt} et~al.(2012){Mandt}, {Waite}, {Teolis}, {Magee}, {Bell},
  {Westlake}, {Nixon}, {Mousis}, and {Lunine}}]{2012ApJ...749..160M}
{Mandt}, K.~E., {Waite}, J.~H., {Teolis}, B., {Magee}, B.~A., {Bell}, J.,
  {Westlake}, J.~H., {Nixon}, C.~A., {Mousis}, O., {Lunine}, J.~I., Apr. 2012.
  {The $^{12}$C/$^{13}$C Ratio on Titan from Cassini INMS Measurements and
  Implications for the Evolution of Methane}. Astrophysical Journal 749, 160.

\bibitem[{{Mastrogiuseppe} et~al.(2014){Mastrogiuseppe}, {Poggiali}, {Hayes},
  {Lorenz}, {Lunine}, {Picardi}, {Seu}, {Flamini}, {Mitri}, {Notarnicola},
  {Paillou}, and {Zebker}}]{LigeiaDepth}
{Mastrogiuseppe}, M., {Poggiali}, V., {Hayes}, A., {Lorenz}, R., {Lunine}, J.,
  {Picardi}, G., {Seu}, R., {Flamini}, E., {Mitri}, G., {Notarnicola}, C.,
  {Paillou}, P., {Zebker}, H., Mar. 2014. {The bathymetry of a Titan sea}. GRL
  41, 1432--1437.

\bibitem[{{McCord} et~al.(2008){McCord}, {Hayne}, {Combe}, {Hansen}, {Barnes},
  {Rodriguez}, {Le Mou{\'e}lic}, {Baines}, {Buratti}, {Sotin}, {Nicholson},
  {Jaumann}, {Nelson}, and {the Cassini VIMS Team}}]{McCord2008}
{McCord}, T.~B., {Hayne}, P., {Combe}, J.-P., {Hansen}, G.~B., {Barnes}, J.~W.,
  {Rodriguez}, S., {Le Mou{\'e}lic}, S., {Baines}, E.~K.~H., {Buratti}, B.~J.,
  {Sotin}, C., {Nicholson}, P., {Jaumann}, R., {Nelson}, R., {the Cassini VIMS
  Team}, Mar. 2008. {Titan's surface: Search for spectral diversity and
  composition using the Cassini VIMS investigation}. Icarus 194, 212--242.

\bibitem[{{Mitchell}(2008)}]{2008JGRE..113.8015M}
{Mitchell}, J.~L., Aug. 2008. {The drying of Titan's dunes: Titan's methane
  hydrology and its impact on atmospheric circulation}. Journal of Geophysical
  Research (Planets) 113, 8015.

\bibitem[{{Mitchell} et~al.(2009){Mitchell}, {Pierrehumbert}, {Frierson}, and
  {Caballero}}]{2009Icar..203..250M}
{Mitchell}, J.~L., {Pierrehumbert}, R.~T., {Frierson}, D.~M.~W., {Caballero},
  R., Sep. 2009. {The impact of methane thermodynamics on seasonal convection
  and circulation in a model Titan atmosphere}. Icarus 203, 250--264.

\bibitem[{{Mitri} et~al.(2007){Mitri}, {Showman}, {Lunine}, and
  {Lorenz}}]{2007Icar..186..385M}
{Mitri}, G., {Showman}, A.~P., {Lunine}, J.~I., {Lorenz}, R.~D., Feb. 2007.
  {Hydrocarbon lakes on Titan}. Icarus 186, 385--394.

\bibitem[{{Moore} and {Howard}(2010)}]{Moore.Tui.Hotei.Lakes}
{Moore}, J.~M., {Howard}, A.~D., Nov. 2010. {Are the basins of Titan's Hotei
  Regio and Tui Regio sites of former low latitude seas?} GRL 37, L22205.

\bibitem[{{Niemann} et~al.(2005){Niemann}, {Atreya}, {Bauer}, {Carignan},
  {Demick}, {Frost}, {Gautier}, {Haberman}, {Harpold}, {Hunten}, {Israel},
  {Lunine}, {Kasprzak}, {Owen}, {Paulkovich}, {Raulin}, {Raaen}, and
  {Way}}]{2005Natur.438..779N}
{Niemann}, H.~B., {Atreya}, S.~K., {Bauer}, S.~J., {Carignan}, G.~R., {Demick},
  J.~E., {Frost}, R.~L., {Gautier}, D., {Haberman}, J.~A., {Harpold}, D.~N.,
  {Hunten}, D.~M., {Israel}, G., {Lunine}, J.~I., {Kasprzak}, W.~T., {Owen},
  T.~C., {Paulkovich}, M., {Raulin}, F., {Raaen}, E., {Way}, S.~H., Dec. 2005.
  {The abundances of constituents of Titan's atmosphere from the GCMS
  instrument on the Huygens probe}. Nature 438, 779--784.

\bibitem[{{Nixon} et~al.(2012){Nixon}, {Temelso}, {Vinatier}, {Teanby},
  {B{\'e}zard}, {Achterberg}, {Mandt}, {Sherrill}, {Irwin}, {Jennings},
  {Romani}, {Coustenis}, and {Flasar}}]{2012ApJ...749..159N}
{Nixon}, C.~A., {Temelso}, B., {Vinatier}, S., {Teanby}, N.~A., {B{\'e}zard},
  B., {Achterberg}, R.~K., {Mandt}, K.~E., {Sherrill}, C.~D., {Irwin},
  P.~G.~J., {Jennings}, D.~E., {Romani}, P.~N., {Coustenis}, A., {Flasar},
  F.~M., Apr. 2012. {Isotopic Ratios in Titan's Methane: Measurements and
  Modeling}. Astrophysical Journal 749, 159.

\bibitem[{{Radebaugh} et~al.(2007){Radebaugh}, {Lorenz}, {Kirk}, {Lunine},
  {Stofan}, {Lopes}, {Wall}, and {the Cassini Radar Team}}]{RADARmountains}
{Radebaugh}, J., {Lorenz}, R.~D., {Kirk}, R.~L., {Lunine}, J.~I., {Stofan},
  E.~R., {Lopes}, R.~M.~C., {Wall}, S.~D., {the Cassini Radar Team}, Dec. 2007.
  {Mountains on Titan observed by Cassini Radar}. Icarus 192, 77--91.

\bibitem[{{Radebaugh} et~al.(2008){Radebaugh}, {Lorenz}, {Lunine}, {Wall},
  {Boubin}, {Reffet}, {Kirk}, {Lopes}, {Stofan}, {Soderblom}, {Allison},
  {Janssen}, {Paillou}, {Callahan}, {Spencer}, and {The Cassini Radar
  Team}}]{2008Icar..194..690R}
{Radebaugh}, J., {Lorenz}, R.~D., {Lunine}, J.~I., {Wall}, S.~D., {Boubin}, G.,
  {Reffet}, E., {Kirk}, R.~L., {Lopes}, R.~M., {Stofan}, E.~R., {Soderblom},
  L., {Allison}, M., {Janssen}, M., {Paillou}, P., {Callahan}, P., {Spencer},
  C., {The Cassini Radar Team}, Apr. 2008. {Dunes on Titan observed by Cassini
  Radar}. Icarus 194, 690--703.

\bibitem[{{Rannou} et~al.(2004){Rannou}, {Hourdin}, {McKay}, and
  {Luz}}]{2004Icar..170..443R}
{Rannou}, P., {Hourdin}, F., {McKay}, C.~P., {Luz}, D., Aug. 2004. {A coupled
  dynamics-microphysics model of Titan's atmosphere}. Icarus 170, 443--462.

\bibitem[{{Rannou} et~al.(2003){Rannou}, {McKay}, and
  {Lorenz}}]{2003P&SS...51..963R}
{Rannou}, P., {McKay}, C.~P., {Lorenz}, R.~D., Dec. 2003. {A model of Titan's
  haze of fractal aerosols constrained by multiple observations}. P\&SS 51,
  963--976.

\bibitem[{{Rannou} et~al.(2006){Rannou}, {Montmessin}, {Hourdin}, and
  {Lebonnois}}]{2006Sci...311..201R}
{Rannou}, P., {Montmessin}, F., {Hourdin}, F., {Lebonnois}, S., Jan. 2006. {The
  Latitudinal Distribution of Clouds on Titan}. Science 311, 201--205.

\bibitem[{{Ribas} et~al.(2005){Ribas}, {Guinan}, {Gudel}, and
  {Audard}}]{Ribas2005}
{Ribas}, I., {Guinan}, E.~F., {Gudel}, M., {Audard}, M., 2005. Evolution of the
  solar activity over time and effects on planetary atmospheres. i. high-energy
  irradiances (1-1700 a). The Astrophysical Journal 622~(1), 680.
\newline\urlprefix\url{http://stacks.iop.org/0004-637X/622/i=1/a=680}

\bibitem[{{Rodriguez} et~al.(2014){Rodriguez}, {Garcia}, {Lucas}, {Appéré}, {Le
  Gall}, {Reffet}, {Le Corre}, {Le Mouélic}, {Cornet}, {Courrech du Pont},
  {Narteau}, {Bourgeois}, {Radebaugh}, {Arnold}, {Barnes}, {Stephan},
  {Jaumann}, {Sotin}, {Brown}, {Lorenz}, and {Turtle}}]{Rodriguez2014}
{Rodriguez}, S., {Garcia}, A., {Lucas}, A., {Appéré}, T., {Le Gall}, A.,
  {Reffet}, E., {Le Corre}, L., {Le Mouélic}, S., {Cornet}, T., {Courrech du
  Pont}, S., {Narteau}, C., {Bourgeois}, O., {Radebaugh}, J., {Arnold}, K.,
  {Barnes}, J., {Stephan}, K., {Jaumann}, R., {Sotin}, C., {Brown}, R.,
  {Lorenz}, R., {Turtle}, E., 2014. {Global mapping and characterization of
  Titan's dune fields with Cassini: correlation between RADAR and VIMS
  observations }. Icarus.
\newline\urlprefix\url{http://www.sciencedirect.com/science/article/pii/S0019103513004892}

\bibitem[{{Rodriguez} et~al.(2011){Rodriguez}, {Le Mou{\'e}lic}, {Rannou},
  {Sotin}, {Brown}, {Barnes}, {Griffith}, {Burgalat}, {Baines}, {Buratti},
  {Clark}, and {Nicholson}}]{2011Icar..216...89R}
{Rodriguez}, S., {Le Mou{\'e}lic}, S., {Rannou}, P., {Sotin}, C., {Brown},
  R.~H., {Barnes}, J.~W., {Griffith}, C.~A., {Burgalat}, J., {Baines}, K.~H.,
  {Buratti}, B.~J., {Clark}, R.~N., {Nicholson}, P.~D., Nov. 2011. {Titan's
  cloud seasonal activity from winter to spring with Cassini/VIMS}. Icarus 216,
  89--110.

\bibitem[{{Rodriguez} et~al.(2009){Rodriguez}, {Le Mou{\'e}lic}, {Rannou},
  {Tobie}, {Baines}, {Barnes}, {Griffith}, {Hirtzig}, {Pitman}, {Sotin},
  {Brown}, {Buratti}, {Clark}, and {Nicholson}}]{2009Natur.459..678R}
{Rodriguez}, S., {Le Mou{\'e}lic}, S., {Rannou}, P., {Tobie}, G., {Baines},
  K.~H., {Barnes}, J.~W., {Griffith}, C.~A., {Hirtzig}, M., {Pitman}, K.~M.,
  {Sotin}, C., {Brown}, R.~H., {Buratti}, B.~J., {Clark}, R.~N., {Nicholson},
  P.~D., Jun. 2009. {Global circulation as the main source of cloud activity on
  Titan}. Nature 459, 678--682.

\bibitem[{{Rodriguez} et~al.(2006){Rodriguez}, {Le Mou{\'e}lic}, {Sotin},
  {Cl{\'e}net}, {Clark}, {Buratti}, {Brown}, {McCord}, {Nicholson}, {Baines},
  and {the VIMS Science Team}}]{Rodriguez.landingsite}
{Rodriguez}, S., {Le Mou{\'e}lic}, S., {Sotin}, C., {Cl{\'e}net}, H., {Clark},
  R.~N., {Buratti}, B., {Brown}, R.~H., {McCord}, T.~B., {Nicholson}, P.~D.,
  {Baines}, K.~H., {the VIMS Science Team}, Dec. 2006. {Cassini/VIMS
  hyperspectral observations of the HUYGENS landing site on Titan}. P\&SS 54,
  1510--1523.

\bibitem[{{Roe}(2012)}]{Roe2012}
{Roe}, H.~G., May 2012. {Titan's Methane Weather}. Annual Review of Earth and
  Planetary Sciences 40, 355--382.

\bibitem[{{Schneider} et~al.(2012){Schneider}, {Graves}, {Schaller}, and
  {Brown}}]{2012Natur.481...58S}
{Schneider}, T., {Graves}, S.~D.~B., {Schaller}, E.~L., {Brown}, M.~E., Jan.
  2012. {Polar methane accumulation and rainstorms on Titan from simulations of
  the methane cycle}. Nature 481, 58--61.

\bibitem[{{Soderblom} et~al.(2009){Soderblom}, {Brown}, {Soderblom}, {Barnes},
  {Kirk}, {Sotin}, {Jaumann}, {MacKinnon}, {Mackowski}, {Baines}, {Buratti},
  {Clark}, and {Nicholson}}]{2009Icar..204..610S}
{Soderblom}, L.~A., {Brown}, R.~H., {Soderblom}, J.~M., {Barnes}, J.~W.,
  {Kirk}, R.~L., {Sotin}, C., {Jaumann}, R., {MacKinnon}, D.~J., {Mackowski},
  D.~W., {Baines}, K.~H., {Buratti}, B.~J., {Clark}, R.~N., {Nicholson}, P.~D.,
  Dec. 2009. {The geology of Hotei Regio, Titan: Correlation of Cassini VIMS
  and RADAR}. Icarus 204, 610--618.

\bibitem[{{Soderblom} et~al.(2007){Soderblom}, {Kirk}, {Lunine}, {Anderson},
  {Baines}, {Barnes}, {Barrett}, {Brown}, {Buratti}, {Clark}, {Cruikshank},
  {Elachi}, {Janssen}, {Jaumann}, {Karkoschka}, {Mou{\'e}lic}, {Lopes},
  {Lorenz}, {McCord}, {Nicholson}, {Radebaugh}, {Rizk}, {Sotin}, {Stofan},
  {Sucharski}, {Tomasko}, and {Wall}}]{2007P&SS...55.2025S}
{Soderblom}, L.~A., {Kirk}, R.~L., {Lunine}, J.~I., {Anderson}, J.~A.,
  {Baines}, K.~H., {Barnes}, J.~W., {Barrett}, J.~M., {Brown}, R.~H.,
  {Buratti}, B.~J., {Clark}, R.~N., {Cruikshank}, D.~P., {Elachi}, C.,
  {Janssen}, M.~A., {Jaumann}, R., {Karkoschka}, E., {Mou{\'e}lic}, S.~L.,
  {Lopes}, R.~M., {Lorenz}, R.~D., {McCord}, T.~B., {Nicholson}, P.~D.,
  {Radebaugh}, J., {Rizk}, B., {Sotin}, C., {Stofan}, E.~R., {Sucharski},
  T.~L., {Tomasko}, M.~G., {Wall}, S.~D., Nov. 2007. {Correlations between
  Cassini VIMS spectra and RADAR SAR images: Implications for Titan's surface
  composition and the character of the Huygens Probe Landing Site}. P\&SS 55,
  2025--2036.

\bibitem[{{Solomonidou} et~al.(submitted){Solomonidou}, {Hirtzig}, {Coustenis},
  {Bratsolis}, {Le Mouelic}, {Rodriguez}, {Stephan}, {Drossart}, {Sotin},
  {Jaumann}, {Brown}, {Kyriakopoulos}, {Lopes}, {Bampasidis},
  {Stamatelopoulou-Seymour}, and {Moussas}}]{SolomonidouPCP}
{Solomonidou}, A., {Hirtzig}, M., {Coustenis}, E.~B., {Bratsolis}, E., {Le
  Mouelic}, S., {Rodriguez}, S., {Stephan}, K., {Drossart}, P., {Sotin}, C.,
  {Jaumann}, R., {Brown}, R.~H., {Kyriakopoulos}, K., {Lopes}, R. M.~C.,
  {Bampasidis}, G., {Stamatelopoulou-Seymour}, K., {Moussas}, X., submitted.
  {Surface albedo spectral properties of geologically interesting areas on
  Titan}. JGR Planets.

\bibitem[{{Sotin} et~al.(2012){Sotin}, {Lawrence}, {Reinhardt}, {Barnes},
  {Brown}, {Hayes}, {Le Mou{\'e}lic}, {Rodriguez}, {Soderblom}, {Soderblom},
  {Baines}, {Buratti}, {Clark}, {Jaumann}, {Nicholson}, and
  {Stephan}}]{2012Icar..221..768S}
{Sotin}, C., {Lawrence}, K.~J., {Reinhardt}, B., {Barnes}, J.~W., {Brown},
  R.~H., {Hayes}, A.~G., {Le Mou{\'e}lic}, S., {Rodriguez}, S., {Soderblom},
  J.~M., {Soderblom}, L.~A., {Baines}, K.~H., {Buratti}, B.~J., {Clark}, R.~N.,
  {Jaumann}, R., {Nicholson}, P.~D., {Stephan}, K., Nov. 2012. {Observations of
  Titan's Northern lakes at 5 {$\mu$}m: Implications for the organic cycle and
  geology}. Icarus 221, 768--786.

\bibitem[{{Sotin} et~al.(2014){Sotin}, {Seignovert}, {Lawrence}, {MacKenzie},
  {Barnes}, and {Brown}}]{SotinEquipotential}
{Sotin}, C., {Seignovert}, B., {Lawrence}, K.~J., {MacKenzie}, S.~M., {Barnes},
  J.~W., {Brown}, R.~H., 2014. {Titan's geoid and hydrology: implications for
  Titan's geological evolution}. in prep.

\bibitem[{{Stiles} et~al.(2009){Stiles}, {Hensley}, {Gim}, {Bates}, {Kirk},
  {Hayes}, {Radebaugh}, {Lorenz}, {Mitchell}, {Callahan}, {Zebker}, {Johnson},
  {Wall}, {Lunine}, {Wood}, {Janssen}, {Pelletier}, {West}, {Veeramacheneni},
  and {Cassini RADAR Team}}]{2009Icar..202..584S}
{Stiles}, B.~W., {Hensley}, S., {Gim}, Y., {Bates}, D.~M., {Kirk}, R.~L.,
  {Hayes}, A., {Radebaugh}, J., {Lorenz}, R.~D., {Mitchell}, K.~L., {Callahan},
  P.~S., {Zebker}, H., {Johnson}, W.~T.~K., {Wall}, S.~D., {Lunine}, J.~I.,
  {Wood}, C.~A., {Janssen}, M., {Pelletier}, F., {West}, R.~D.,
  {Veeramacheneni}, C., {Cassini RADAR Team}, Aug. 2009. {Determining Titan
  surface topography from Cassini SAR data}. Icarus 202, 584--598.

\bibitem[{{Stofan} et~al.(2012){Stofan}, {Aharonson}, {Hayes}, {Kirk}, {Lopes},
  {Lorenz}, {Lucas}, {Lunine}, {Malaska}, {Radebaugh}, {Stiles}, {Turtle},
  {Wall}, {Wood}, and {Cassini Radar Team}}]{2012DPS....4420108S}
{Stofan}, E.~R., {Aharonson}, O., {Hayes}, A.~G., {Kirk}, R., {Lopes}, R.,
  {Lorenz}, R.~D., {Lucas}, A., {Lunine}, J.~I., {Malaska}, M., {Radebaugh},
  J., {Stiles}, B.~W., {Turtle}, E.~P., {Wall}, S.~D., {Wood}, C.~A., {Cassini
  Radar Team}, Oct. 2012. {Searching for the Remnants of Southern Seas: Cassini
  Observations of the South Pole of Titan}. In: AAS/Division for Planetary
  Sciences Meeting Abstracts. Vol.~44 of AAS/Division for Planetary Sciences
  Meeting Abstracts. p. \#201.08.

\bibitem[{{Stofan} et~al.(2007){Stofan}, {Elachi}, {Lunine}, {Lorenz},
  {Stiles}, {Mitchell}, {Ostro}, {Soderblom}, {Wood}, {Zebker}, {Wall},
  {Janssen}, {Kirk}, {Lopes}, {Paganelli}, {Radebaugh}, {Wye}, {Anderson},
  {Allison}, {Boehmer}, {Callahan}, {Encrenaz}, {Flamini}, {Francescetti},
  {Gim}, {Hamilton}, {Hensley}, {Johnson}, {Kelleher}, {Muhleman}, {Paillou},
  {Picardi}, {Posa}, {Roth}, {Seu}, {Shaffer}, {Vetrella}, and
  {West}}]{2007Natur.445...61S}
{Stofan}, E.~R., {Elachi}, C., {Lunine}, J.~I., {Lorenz}, R.~D., {Stiles}, B.,
  {Mitchell}, K.~L., {Ostro}, S., {Soderblom}, L., {Wood}, C., {Zebker}, H.,
  {Wall}, S., {Janssen}, M., {Kirk}, R., {Lopes}, R., {Paganelli}, F.,
  {Radebaugh}, J., {Wye}, L., {Anderson}, Y., {Allison}, M., {Boehmer}, R.,
  {Callahan}, P., {Encrenaz}, P., {Flamini}, E., {Francescetti}, G., {Gim}, Y.,
  {Hamilton}, G., {Hensley}, S., {Johnson}, W.~T.~K., {Kelleher}, K.,
  {Muhleman}, D., {Paillou}, P., {Picardi}, G., {Posa}, F., {Roth}, L., {Seu},
  R., {Shaffer}, S., {Vetrella}, S., {West}, R., Jan. 2007. {The lakes of
  Titan}. Nature 445, 61--64.

\bibitem[{{Teanby} et~al.(2012){Teanby}, {Irwin}, {Nixon}, {de Kok},
  {Vinatier}, {Coustenis}, {Sefton-Nash}, {Calcutt}, and
  {Flasar}}]{2012Natur.491..732T}
{Teanby}, N.~A., {Irwin}, P.~G.~J., {Nixon}, C.~A., {de Kok}, R., {Vinatier},
  S., {Coustenis}, A., {Sefton-Nash}, E., {Calcutt}, S.~B., {Flasar}, F.~M.,
  Nov. 2012. {Active upper-atmosphere chemistry and dynamics from polar
  circulation reversal on Titan}. Nature 491, 732--735.

\bibitem[{{Tokano}(2005)}]{Tokano05}
{Tokano}, T., Jan 2005. Meteorological assessment of the surface temperatures
  on titan: constraints on the surface type. Icarus 173, 222--242.

\bibitem[{{Tokano}(2009)}]{2009Icar..204..619T}
{Tokano}, T., Dec. 2009. {Impact of seas/lakes on polar meteorology of Titan:
  Simulation by a coupled GCM-Sea model}. Icarus 204, 619--636.

\bibitem[{{Tomasko} et~al.(2005){Tomasko}, {Archinal}, {Becker}, {B{\'e}zard},
  {Bushroe}, {Combes}, {Cook}, {Coustenis}, {de Bergh}, {Dafoe}, {Doose},
  {Dout{\'e}}, {Eibl}, {Engel}, {Gliem}, {Grieger}, {Holso}, {Howington-Kraus},
  {Karkoschka}, {Keller}, {Kirk}, {Kramm}, {K{\"u}ppers}, {Lanagan},
  {Lellouch}, {Lemmon}, {Lunine}, {McFarlane}, {Moores}, {Prout}, {Rizk},
  {Rosiek}, {Rueffer}, {Schr{\"o}der}, {Schmitt}, {See}, {Smith}, {Soderblom},
  {Thomas}, and {West}}]{2005Natur.438..765T}
{Tomasko}, M.~G., {Archinal}, B., {Becker}, T., {B{\'e}zard}, B., {Bushroe},
  M., {Combes}, M., {Cook}, D., {Coustenis}, A., {de Bergh}, C., {Dafoe},
  L.~E., {Doose}, L., {Dout{\'e}}, S., {Eibl}, A., {Engel}, S., {Gliem}, F.,
  {Grieger}, B., {Holso}, K., {Howington-Kraus}, E., {Karkoschka}, E.,
  {Keller}, H.~U., {Kirk}, R., {Kramm}, R., {K{\"u}ppers}, M., {Lanagan}, P.,
  {Lellouch}, E., {Lemmon}, M., {Lunine}, J., {McFarlane}, E., {Moores}, J.,
  {Prout}, G.~M., {Rizk}, B., {Rosiek}, M., {Rueffer}, P., {Schr{\"o}der},
  S.~E., {Schmitt}, B., {See}, C., {Smith}, P., {Soderblom}, L., {Thomas}, N.,
  {West}, R., Dec. 2005. {Rain, winds and haze during the Huygens probe's
  descent to Titan's surface}. Nature 438, 765--778.

\bibitem[{{Turtle} et~al.(2011{\natexlab{a}}){Turtle}, {Perry}, {Hayes},
  {Lorenz}, {Barnes}, {McEwen}, {West}, {Del Genio}, {Barbara}, {Lunine},
  {Schaller}, {Ray}, {Lopes}, and {Stofan}}]{2011Sci...331.1414T}
{Turtle}, E.~P., {Perry}, J.~E., {Hayes}, A.~G., {Lorenz}, R.~D., {Barnes},
  J.~W., {McEwen}, A.~S., {West}, R.~A., {Del Genio}, A.~D., {Barbara}, J.~M.,
  {Lunine}, J.~I., {Schaller}, E.~L., {Ray}, T.~L., {Lopes}, R.~M.~C.,
  {Stofan}, E.~R., Mar. 2011{\natexlab{a}}. {Rapid and Extensive Surface
  Changes Near Titan's Equator: Evidence of April Showers}. Science 331,
  1414--1417.

\bibitem[{{Turtle} et~al.(2011{\natexlab{b}}){Turtle}, {Perry}, {Hayes}, and
  {McEwen}}]{Ontario.dries.up}
{Turtle}, E.~P., {Perry}, J.~E., {Hayes}, A.~G., {McEwen}, A.~S., Apr.
  2011{\natexlab{b}}. {Shoreline retreat at Titan's Ontario Lacus and Arrakis
  Planitia from Cassini Imaging Science Subsystem observations}. Icarus 212,
  957--959.

\bibitem[{{Turtle} et~al.(2009){Turtle}, {Perry}, {McEwen}, {Del Genio},
  {Barbara}, {West}, {Dawson}, and {Porco}}]{2009GeoRL..3602204T}
{Turtle}, E.~P., {Perry}, J.~E., {McEwen}, A.~S., {Del Genio}, A.~D.,
  {Barbara}, J., {West}, R.~A., {Dawson}, D.~D., {Porco}, C.~C., Jan. 2009.
  {Cassini imaging of Titan's high-latitude lakes, clouds, and south-polar
  surface changes}. Geophysical Research Letters 36, L2204.

\bibitem[{{Vixie} et~al.(submitted){Vixie}, {Barnes}, {Jackson}, {Rodriguez},
  {Le Mouelic}, {Sotin}, and {Wilson}}]{VixieLakes}
{Vixie}, G., {Barnes}, J., {Jackson}, B., {Rodriguez}, S., {Le Mouelic}, S.,
  {Sotin}, C., {Wilson}, P., submitted. {Temperate Lakes on Titan}. Icarus.

\bibitem[{{Wahlund} et~al.(2009){Wahlund}, {Galand}, {M{\"u}ller-Wodarg},
  {Cui}, {Yelle}, {Crary}, {Mandt}, {Magee}, {Waite}, {Young}, {Coates},
  {Garnier}, {{\AA}gren}, {Andr{\'e}}, {Eriksson}, {Cravens}, {Vuitton},
  {Gurnett}, and {Kurth}}]{2009P&SS...57.1857W}
{Wahlund}, J.-E., {Galand}, M., {M{\"u}ller-Wodarg}, I., {Cui}, J., {Yelle},
  R.~V., {Crary}, F.~J., {Mandt}, K., {Magee}, B., {Waite}, J.~H., {Young},
  D.~T., {Coates}, A.~J., {Garnier}, P., {{\AA}gren}, K., {Andr{\'e}}, M.,
  {Eriksson}, A.~I., {Cravens}, T.~E., {Vuitton}, V., {Gurnett}, D.~A.,
  {Kurth}, W.~S., Dec. 2009. {On the amount of heavy molecular ions in Titan's
  ionosphere}. P\&SS 57, 1857--1865.

\bibitem[{{Wall} et~al.(2010){Wall}, {Hayes}, {Bristow}, {Lorenz}, {Stofan},
  {Lunine}, {Le Gall}, {Janssen}, {Lopes}, {Wye}, {Soderblom}, {Paillou},
  {Aharonson}, {Zebker}, {Farr}, {Mitri}, {Kirk}, {Mitchell}, {Notarnicola},
  {Casarano}, and {Ventura}}]{2010GeoRL..3705202W}
{Wall}, S., {Hayes}, A., {Bristow}, C., {Lorenz}, R., {Stofan}, E., {Lunine},
  J., {Le Gall}, A., {Janssen}, M., {Lopes}, R., {Wye}, L., {Soderblom}, L.,
  {Paillou}, P., {Aharonson}, O., {Zebker}, H., {Farr}, T., {Mitri}, G.,
  {Kirk}, R., {Mitchell}, K., {Notarnicola}, C., {Casarano}, D., {Ventura}, B.,
  Mar. 2010. {Active shoreline of Ontario Lacus, Titan: A morphological study
  of the lake and its surroundings}. GRL 37, L5202.

\bibitem[{{Wasiak} et~al.(2013){Wasiak}, {Androes}, {Blackburn}, {Tullis},
  {Dixon}, and {Chevrier}}]{Wasiak2013}
{Wasiak}, F.~C., {Androes}, D., {Blackburn}, D.~G., {Tullis}, J.~A., {Dixon},
  J., {Chevrier}, V.~F., Aug. 2013. {A geological characterization of Ligeia
  Mare in the northern polar region of Titan}. P\&SS 84, 141--147.

\bibitem[{{Wood} et~al.(2013){Wood}, {Stofan}, {Hayes}, {Kirk}, {Lunine},
  {Radebaugh}, and {Malaska}}]{2013LPI....44.1764W}
{Wood}, C.~A., {Stofan}, E.~R., {Hayes}, A.~G., {Kirk}, R.~K., {Lunine}, J.~I.,
  {Radebaugh}, J., {Malaska}, M., Mar. 2013. {Morphological Evidence for Former
  Seas Near Titan's South Pole}. In: Lunar and Planetary Institute Science
  Conference Abstracts. Vol.~44 of Lunar and Planetary Institute Science
  Conference Abstracts. p. 1764.

\end{thebibliography}

\end{document}